\documentclass[12pt]{article}

\usepackage{a4wide,epsfig,psfrag,cite}
\usepackage{amsmath,amssymb,scalefnt}
\usepackage{color}
\usepackage{slashed}
\usepackage{subcaption}

\graphicspath{ {figs/} }

\definecolor{darkgreen}{cmyk}{1.0,0,1.0,0.61}
\definecolor{light-gray}{gray}{0.95}

\allowdisplaybreaks

\begin{document}

\title{\vskip-3cm{\baselineskip14pt
    \begin{flushleft}
      \normalsize TTP18-033
  \end{flushleft}}
  \vskip1.5cm
  Wilson coefficients for Higgs boson production and decoupling relations to $\mathcal{O}\left(\alpha_s^4\right)$
}

\author{
  Marvin Gerlach, Florian Herren and Matthias Steinhauser
  \\[1em]
  {\small\it Institut f{\"u}r Theoretische Teilchenphysik}\\
  {\small\it Karlsruhe Institute of Technology (KIT)}\\
  {\small\it 76128 Karlsruhe, Germany}
}

\date{}

\maketitle

\begin{abstract}
An important ingredient for the calculation of Higgs boson properties
in the infinite top quark mass limit is the knowledge of the effective
coupling between the Higgs bosons and gluons, i.e. the Wilson
coefficients $C_H$ and $C_{HH}$ for one and two Higgs bosons,
respectively.  In this work we calculate for the first time $C_{HH}$
to four loops in a direct, diagrammatic way, discussing in detail all
issues arising due to the renormalization of operator products.
Furthermore, we also calculate the Wilson coefficient $C_H$ for the
coupling of a single Higgs boson to gluons as well as all four loop
decoupling relations in QCD with general SU$(N_c)$ colour factors. The latter
are related to $C_H$ and $C_{HH}$ via low-energy theorems.
\end{abstract}

\thispagestyle{empty}

%- }}}

\newpage

%- {{{ Introduction:

\section{Introduction}

In the coming years the production and decay of Higgs bosons will play a
central role in many analyses performed at the Large Hadron Collider (LHC).  A
crucial ingredient is often provided by the matching coefficients which govern
the coupling of Higgs bosons to gluons. The corresponding effective Lagrange
density is valid in the heavy top quark limit which provides a good
approximation for Higgs boson decays to gluons and the total production cross
section of a single Higgs boson. For less inclusive processes the
applicability of the effective theory approach is limited to parts of the phase
space. This is also true for Higgs boson pair production.  In this paper we
perform for the first time a direct calculation of the matching coefficient
for the coupling of one and two Higgs bosons to gluons.  Our results are
expressed in terms of general SU($N_c$) colour factors.  For $N_c=3$ they can be
compared to expressions obtained from indirect methods where the matching
coefficients are obtained from low-energy theorems (LETs).  The essential
ingredient into the LETs is the QCD decoupling constant for the strong
coupling. For this reason we re-visit the calculation of all four-loop
decoupling constants and provide results for a generic SU($N_c$) gauge group.

The remainder of this paper is organized as follows: In Section~\ref{sec:tech}
we fix our notation and introduce the decoupling constants and the effective
Lagrange density for the Higgs-gluon coupling.  In Sections~\ref{sec:dec}
and~\ref{sec:higgs} we present our results for the decoupling relations and
Wilson coefficients, respectively. We discuss in detail the extraction of the
coupling of two Higgs bosons to gluons ($C_{HH}$) and in particular the
subtleties in the matching procedure due to the renormalization of products of
operators.  Our findings are summarized in Section~\ref{sec:conclusions}.  In
the Appendix we collect analytic results for the decoupling constants.

%- }}}
%- {{{ Technicalities:

\section{\label{sec:tech}Technicalities}

For convenience of the reader and to fix our notation we repeat in this
Section the definition of the decoupling constants in QCD and the
Wilson coefficients in the effective Lagrange density describing Higgs-gluon
couplings. For a detailed discussion we refer to Ref.~\cite{Chetyrkin:1997un}.
We will work in the $\overline{\mathrm{MS}}$ scheme throughout this paper,
except for the heavy quark mass which we renormalize both in the
$\overline{\mathrm{MS}}$ and on-shell scheme.
The $\overline{\rm MS}$ counterterms are needed up to four-loop order (see,
e.g., Ref.~\cite{Chetyrkin:2004mf}) and the renormalization constant for the 
$\overline{\rm MS}$ to on-shell conversion for the heavy quark mass to three
loops~\cite{Chetyrkin:1999ys,Chetyrkin:1999qi,Melnikov:2000qh,Marquard:2007uj}. 

%- {{{ Decoupling constants:

\subsection{\label{sub::dec}Decoupling constants}

The bare and renormalized parameters and fields of the QCD Lagrangian are
connected by renormalization constants defined through
\begin{align}
  &&g_s^0 = \mu^{\epsilon}Z_g g_s, &&&m_q^0 = Z_m m_q, &&\xi^0 - 1 = Z_3(\xi-1)\,,
  \nonumber\\
  &&A_\mu^{0,a} = \sqrt{Z_3}A_\mu^a,  &&&\psi^0_q = \sqrt{Z_2}\psi_q,&&c^{0,a} =\sqrt{\tilde{Z}_3}c^a
  \,.
  \label{eq::renconstants}
\end{align}
Here $g_s$ is the QCD gauge coupling with $\alpha_s = g_s^2/(4\pi)$ being the
strong coupling constant, $\mu$ is the renormalization scale,
$D = 4 - 2\epsilon$ the space-time dimension and $\xi$ the gauge parameter
with $\xi = 0$ corresponding to Feynman and $\xi=1$ to Landau gauge. The gluon
field is given by $A^a_\mu$, $\psi_q$ is the quark field of flavour $q$ with
mass $m_q$ and $c^a$ is the ghost field.  Bare quantities are denoted by the
superscript ``0''. The renormalization constants $Z_X$ are needed up to
$\mathcal{O}(\alpha_s^4)$~\cite{Chetyrkin:1997dh,Vermaseren:1997fq,vanRitbergen:1997va,Czakon:2004bu,Chetyrkin:2004mf}
for our purposes.

In the following we assume a strong hierarchy in the quark masses and
integrate out a heavy quark with mass $m_h$ from QCD with $n_f$ active quark
flavours.\footnote{The simultaneous decoupling of two heavy quarks with
  different masses is discussed in Ref.~\cite{Grozin:2011nk} up to three-loop
  order.}  The resulting effective Lagrangian has the same form as the
original QCD Lagrangian. However, it only has $n_l=n_f-1$ active quark
flavours and thus only depends on the light degrees of freedom. The
parameters and fields in the effective $n_l$-flavour and full $n_f$-flavour
theory are related via the so-called (bare) decoupling constants
\begin{align}
  &&g_s^{0\,(n_l)} = \zeta^0_g g^{0\,(n_f)}_s, &&&m_q^{0\,(n_l)} = \zeta^{0\,(n_f)}_m m^0_q,
  &&\xi^{0\,(n_l)} - 1 = \zeta^0_3(\xi^{0\,(n_f)}-1),\nonumber\\
  &&A_\mu^{0\,(n_l)} = \sqrt{\zeta^0_3}A_\mu^{0\,(n_f)},  &&&\psi^{0\,(n_l)}_q = \sqrt{\zeta^0_2}\psi_q^{0\,(n_f)},
  &&c^{0\,(n_l)} = \sqrt{\tilde{\zeta}^0_3}c^{0\,(n_f)}\,,
     \label{eq::deccoeff}
\end{align}
where the superscripts denote the number of active quark flavours.  For
simplicity we refrain to show colour indices for the fields.  The different
decoupling constants $\zeta_X$ contain the radiative effects of the heavy
quark and can be computed in a perturbative series in $\alpha_s$.

One obtains the renormalized decoupling constants by replacing the bare
parameters and fields in Eq.~\eqref{eq::deccoeff} by renormalized counterparts
using Eq.~(\ref{eq::renconstants}).  As an example, consider the gauge coupling
where the renormalized decoupling constant is given by
\begin{align}
  g_s^{(n_l)} = \frac{Z^{(n_f)}_g}{Z^{(n_l)}_g} \zeta^0_g  g^{(n_f)}_s =
  \zeta_g g^{(n_f)}_s
  \,.
  \label{eq::zetags}
\end{align}
Note that $Z^{(n_l)}_g$ depends on $g_s^{(n_l)}$ which has to be transformed
to $g_s^{(n_f)}$ using Eq.~(\ref{eq::zetags}). Thus, it is natural to apply
Eq.~(\ref{eq::zetags}) iteratively to arrive at four
loops. Note that the loop corrections to the renormalization constants only
contain poles whereas the decoupling constants also contain positive powers in
$\epsilon$. Thus, $Z^{(n_l)}_g$ expressed in terms of $g_s^{(n_f)}$ also
contains positive powers in $\epsilon$.

For later convenience we define the decoupling constant for the strong
coupling constant $\alpha_s$ as
\begin{align}
  \zeta_{\alpha_s} = \zeta_g^2\,.
\end{align}

%- }}}
%- {{{ Wilson coefficients for Higgs boson production and decay:

\subsection{\label{sub::wil}Wilson coefficients for Higgs boson production and decay}

In the Standard Model, the coupling of a Higgs boson to gluons is mainly mediated by top
quark loops and thus in the following we have $n_f = 6$ and $n_l = 5$ for the
full and effective theory, respectively.  The effective Lagrange density which
describes the coupling of one or two Higgs boson to gluons is obtained after
integrating out the top quark and is given by
\begin{align}
  \mathcal{L}_\mathrm{eff} = -\frac{H}{v}C_H^0\mathcal{O}_1^0 
  + \frac{1}{2}\left(\frac{H}{v}\right)^2C_{HH}^0\mathcal{O}_1^0\,,
  \label{eq::leff}
\end{align}
where $\mathcal{O}_1 = G_{\mu\nu}^a G^{\mu\nu,a}/4$, with $G_{\mu\nu}^a$ being
the gluon field strength tensor, is the only physical operator one has to
consider. It is defined in the $n_l=5$-flavour effective theory.  The Wilson
coefficients $C_H^0$ and $C_{HH}^0$ comprise the radiative effects of the top
quark, which is in analogy to the decoupling constants introduced in
Eq.~\eqref{eq::deccoeff}.

The renormalization of $\mathcal{O}_1$ has been discussed in detail in
Ref.~\cite{Spiridonov:1984br} (see also Ref.~\cite{Zoller:2016iam}).  In fact,
the renormalization constant $Z_{\mathcal{O}_1}$ can be expressed through the 
QCD beta function through all orders in perturbation theory~\cite{Spiridonov:1984br}
\begin{align}
  Z_{\mathcal{O}_1} = \frac{1}{ 1 -  \beta(\alpha_s^{(5)})/\epsilon }\,,
\end{align}
with
\begin{eqnarray}
  \beta(\alpha_s) &=& - \left(\frac{\alpha_s}{\pi}\right)^2 \sum_{n\ge0} \beta_n
                      \left(\frac{\alpha_s}{\pi}\right)^n \,,\nonumber\\
  \beta_0 &=& \frac{1}{4}\left( \frac{11}{3}C_A - \frac{4}{3}T_Fn_f \right)\,,\nonumber\\
  \beta_1 &=& \frac{1}{16}\left( \frac{34}{3}C_A^2 - \frac{20}{3}C_A T_F n_f -
              4 C_F T_F n_f \right)\,,\nonumber\\
  \beta_2 &=& \frac{1}{64}\left( 
              \frac{2857}{54}C_A^3 - \frac{1415}{27}C^2_A T_F n_f -
              \frac{205}{9}C_A C_F T_F n_f + 2 C_F^2 T_F n_f
  \right.\nonumber\\
          &\phantom{=}& \left.+ \frac{158}{27}C_A T^2_F n^2_f + \frac{44}{9}C_F T^2_F n^2_f\right)\,.
\end{eqnarray}
$Z_{\mathcal{O}_1}$ can be used to obtain the
renormalized Wilson coefficients via the relation
\begin{align}
  C_{X}^0\mathcal{O}_1^0 = \frac{C_{X}^0}{Z_{\mathcal{O}_1}}\,Z_{\mathcal{O}_1}\mathcal{O}_1^0 = C_{X}\mathcal{O}_1\,,
\end{align}
with $X\in\{H,HH\}$.

%- }}}
%- {{{ Low energy theorems:

\subsection{Low energy theorems}

There is a close connection between the decoupling constants from
Subsection~\ref{sub::dec} and the Wilson coefficients from~\ref{sub::wil}
which is established by the so-called LETs.
In \cite{Chetyrkin:1997un} a LET relating $\zeta_{\alpha_s}$ and $C_H$ has been derived
\begin{align}
  C_H = -\frac{m_t}{\zeta_{\alpha_s}}\frac{\partial}{\partial m_t}\zeta_{\alpha_s}
  \label{eq::let_ch}
\end{align}
where we adapted the prefactors to match our conventions.  Beyond three loops
$C_H$ has only been obtained with the help of Eq.~(\ref{eq::let_ch}).  In this
work we perform an explicit calculation of $C_H$ for general SU($N_c$) colour
factors.

Recently, in Ref.~\cite{Spira:2016zna} a LET has been proposed for $C_{HH}$
which reads
\begin{align}
  C_{HH} = 
  \frac{m^2_t}{\zeta_{\alpha_s}}\frac{\partial^2}{\partial 
  m_t^2}\zeta_{\alpha_s}
  - 2\left(\frac{m_t}{\zeta_{\alpha_s}}\frac{\partial}{\partial
  m_t}\zeta_{\alpha_s}\right)^2 
  \,.
  \label{eq::let_chh}
\end{align}
It provides the correct result for $C_{HH}$ at three-loop
order~\cite{Grigo:2014jma};
in Section~\ref{sec:higgs} we perform an explicit calculation of $C_{HH}$ and
show that Eq.~(\ref{eq::let_chh}) also works at four loops.

Note that in QCD $\zeta_{\alpha_s}$ depends on $m_t$ only via logarithms of
the form $\log(\mu^2/m_t^2)$. Thus, it is possible to reconstruct the $m_t$
dependence at $(n+1)$-loop order from the $n$-loop result of
$\zeta_{\alpha_s}$ with the help of renormalization group equations.
Using the LETs this immediately leads to the $(n+1)$-loop results for $C_H$
and $C_{HH}$.

%- }}}
%- {{{ Computational setup:

\subsection{Computational setup}

For our calculation we use a well tested, automated setup, starting with the
generation of Feynman diagrams using \verb|qgraf|~\cite{Nogueira:1991ex}.  The
output is processed by \verb|q2e| and
\verb|exp|~\cite{Harlander:1997zb,Seidensticker:1999bb,q2eexp}, which generate
\verb|FORM|~\cite{Ruijl:2017dtg} code for the amplitudes and map
them onto individual integral families. We then compute the colour factors of the
diagrams using \verb|color|~\cite{vanRitbergen:1998pn} and combine amplitudes
with the same colour factors and integral families to so-called superdiagrams
so that we can process them together.

After processing Lorentz structures and expanding in the external momenta, we
are left with single-scale tensor tadpole integrals. We perform a tensor
decomposition and reduce the remaining, scalar integrals to master integrals,
using \verb|LiteRed| \cite{Lee:2012cn,Lee:2013mka} and
\verb|FIRE5|~\cite{Smirnov:2014hma}. With the help of the \verb|FindRules| command of
\verb|FIRE5| we identify equivalent master integrals from different integral families.

The master integrals are all known to sufficiently high order in $\epsilon$
\cite{Lee:2010hs} (see also \cite{Schroder:2005va,Chetyrkin:2006dh}).  The
missing $\epsilon^3$ term of the integral $J_{6,2}$ (in the notation of
\cite{Lee:2010hs}) was provided in~\cite{Lee:eps3}.

As a cross-check, we also computed the $ggH$ amplitude and the
decoupling constants through three loops using
\verb|MATAD|~\cite{Steinhauser:2000ry}.

%- }}}

%- }}}
%- {{{ Calculation of decoupling constants:

\section{\label{sec:dec}Calculation of decoupling constants}

We aim for the calculation of all QCD decoupling constants up to four-loop
order with general SU($N_c$) colour factors. They are obtained from
$\zeta_3^0$, $\tilde{\zeta}_3^0$, $\zeta_2^0$ and $\zeta_m^0$
as introduced in Eq.~(\ref{eq::deccoeff}) and the decoupling constant
of the ghost-gluon vertex, $\tilde{\zeta}_1^0$.
The decoupling constant for the gauge coupling is then given by
\begin{align}
  \zeta_g^0 = \frac{\tilde{\zeta}^0_1}{\tilde{\zeta}^0_3\sqrt{\zeta^0_3}}\,.
\end{align}
The remaining decoupling constants for the gluon-quark vertex ($\zeta_1^0$),
the three-gluon vertex ($\zeta_{3g}^0$) and four-gluon-vertex ($\zeta_{4g}^0$)
are obtained with the help of the Ward identities
\begin{eqnarray}
  \zeta_1^0 &=& \zeta_g^0 \zeta_2^0 \sqrt{\zeta^0_3}\,,\nonumber\\
  \zeta_{3g}^0 &=& \zeta_g^0 (\zeta^0_3)^{3/2}\,,\nonumber\\
  \zeta_{4g}^0 &=& (\zeta_g^0)^2 (\zeta^0_3)^2\,.
\end{eqnarray}

The bare decoupling constants $\zeta_3^0$, $\tilde{\zeta}_3^0$, $\zeta_2^0$
and $\zeta_m^0$ are obtained from the hard
part of the gluon and ghost vacuum polarizations $\Pi_G(p^2)$ and
$\Pi_c(p^2)$, as well as the vector and scalar parts of the light quark
self-energy $\Sigma_V(p^2)$ and $\Sigma_S(p^2)$ as~\cite{Chetyrkin:1997un}
\begin{align}
  \zeta_3^0 = 1 + \Pi^{0,h}_G(0)~,\nonumber\\
  \tilde{\zeta}_3^0 = 1 + \Pi^{0,h}_c(0)~,\nonumber\\
  \zeta_2^0 = 1 + \Sigma^{0,h}_V(0)~,\nonumber\\
  \zeta_m^0 = \frac{1 - \Sigma^{0,h}_S(0)}{1 + \Sigma^{0,h}_V(0)}\,.
  \label{eq::zeta}
\end{align}
To obtain the decoupling constants we only need the leading term in the
limit $m_h \rightarrow \infty$. Thus, we can Taylor-expand in the external
momenta and set them to zero after factoring out the tree-level tensor
structure. This reduces the integrals to single-scale tadpole integrals.  In
analogy to Eq.~(\ref{eq::zeta}), $\tilde{\zeta}_1^0$ is obtained from the
ghost-gluon vertex
\begin{align}
  \tilde{\zeta}_1^0 = 1 + \Gamma_{G\overline{c}c}(p,q)\Big|_{p,q\to 0}\,.
  \label{eq::zeta1til}
\end{align}
where $p$ and $q$ are the four-momenta of the ghost and gluon, respectively. After projecting out
the tree-level contribution both $p$ and $q$ are set to zero.

\begin{table}[t]
  \begin{center}
    \begin{tabular}[t]{c||r|r|r|r}
      \# loops & 1 & 2 & 3 & 4 \\
      \hline
      $\Pi^{0,h}_G$               & 1    &   7 & 189 &  6\,245 \\
      $\Pi^{0,h}_c$               & ---  &   1 &  25 &  765 \\
      $\Sigma^{0,h}_{V/S}$        & ---  &   1 &  25 &  765 \\
      $\Gamma_{G\overline{c}c}$ & ---  &   5 &  228&  10\,118 \\
    \end{tabular}
    \caption{\label{tbl::numdia_zeta}Number of diagrams needed for computing
      the decoupling constants up to four loops.}
  \end{center}
\end{table}

\begin{figure}[t]
  \begin{center}
    \includegraphics[width=0.22\textwidth]{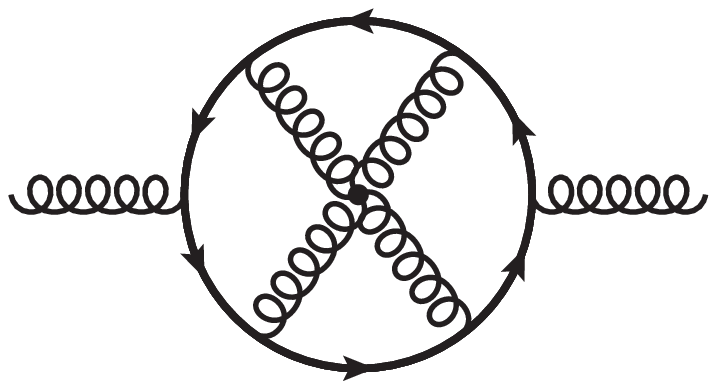}
    \includegraphics[width=0.22\textwidth]{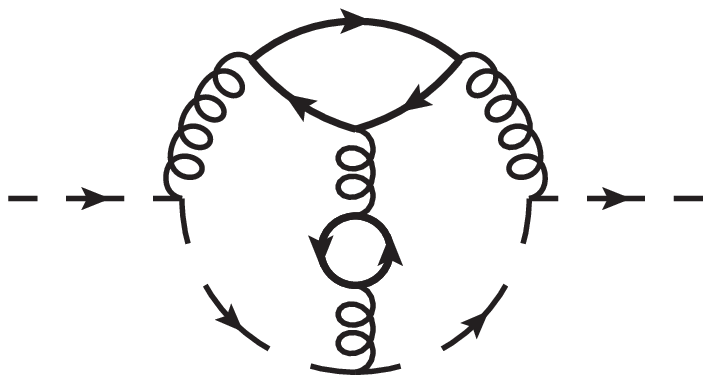}
    \includegraphics[width=0.22\textwidth]{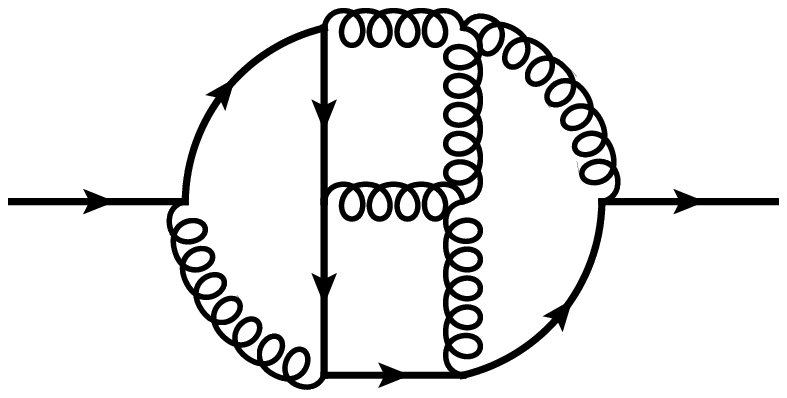}
    \includegraphics[width=0.22\textwidth]{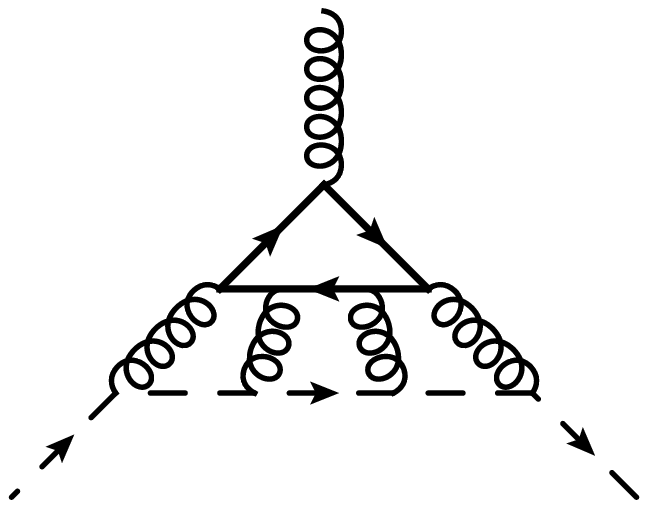}
  \end{center}
  \caption{\label{fig::zeta} Sample four-loop diagrams contributing to the
    decoupling constants defined in Eqs.~(\ref{eq::zeta})
    and~(\ref{eq::zeta1til}). Solid, curly and dashed lines refer to
    fermions, gluons and ghosts, respectively.}
\end{figure}

In Tab.~\ref{tbl::numdia_zeta} we present the number of diagrams generated by
\verb|qgraf| for the individual Green functions. Sample four-loop Feynman diagrams
are shown in Fig.~\ref{fig::zeta}.

We perform the calculation keeping the full dependence on the gauge parameter
$\xi$ which drops out for $\zeta_{\alpha_s}^0$ and $\zeta_m^0$, as expected on
general grounds. All other decoupling constants have an explicit $\xi$
dependence. At three-loop order our results agree with those of
Ref.~\cite{Chetyrkin:1997un} and at four loops we reproduce the results for
$\zeta_{\alpha_s}$ from Refs.~\cite{Schroder:2005hy,Chetyrkin:2005ia} and
$\zeta_m$ from~\cite{Liu:2015fxa} after specifying $N_c=3$.

The decoupling constants $\zeta_{\alpha_s}$ and $\zeta_m$ as well as the
leading terms of the others can be found in Appendix~\ref{app:dec}.  We
provide the results for all renormalized decoupling constants in computer
readable form in the ancillary files~\cite{progdata}. For convenience 
we offer several options concerning the renormalization scheme of
the heavy quark ($\overline{\rm MS}$ vs. on-shell) and $\alpha_s$
($n_f$ vs. $n_l$ active flavours).

%- }}}
%- {{{ Direct calculation of matching coefficients:

\section{\label{sec:higgs}Direct calculation of matching coefficients}

This section is devoted to the direct calculation of $C_H$ and $C_{HH}$
defined in the
effective Lagrange density in Eq.~(\ref{eq::leff}).  Two-loop results for $C_H$ are
known since the beginning of the eighties~\cite{Inami:1982xt,Djouadi:1991tka}
and at three-loop order $C_H$ has been obtained for the first time from a
direct calculation of the Higgs-gluon vertex in the large-$m_t$ limit in
Ref.~\cite{Chetyrkin:1997iv} (see also Ref.~\cite{Steinhauser:2002rq}). Later
the result has been confirmed with the help of a LET derived in
Ref.~\cite{Chetyrkin:1997iv} (see also Ref.~\cite{Kramer:1996iq}).  Using the
three-loop decoupling constant for $\alpha_s$, the LET in combination with the
four-loop beta function~\cite{vanRitbergen:1997va,Czakon:2004bu} even leads to
the four-loop result for $C_H$. The same reasoning has been applied in
Refs.~\cite{Schroder:2005hy,Chetyrkin:2005ia,Chetyrkin:2016uhw} to obtain the
five-loop prediction for $C_H$, where an important input is provided by (the
fermionic part of) the five-loop beta function which has been computed in
Refs.~\cite{Baikov:2016tgj,Herzog:2017ohr,Luthe:2017ttg}.  To date there is no
direct calculation of $C_H$ at four loops.

For $C_{HH}$ the situation is as follows: at one- and two-loop order $C_{HH}$
and $C_H$ agree. At three-loop order a direct calculation has been performed
in Ref.~\cite{Grigo:2014jma} by matching the full to the effective theory
in Eq.~(\ref{eq::leff}). The result has been confirmed via the LET from
Ref.~\cite{Spira:2016zna}, which can be used to derive the
four-loop result for $C_{HH}$. It is one of the main aims of this paper
to perform a direct calculation of $C_{HH}$ and to confront it with the LET
result. 

In the following we use the notation for the matching equations introduced in
Ref.~\cite{Grigo:2014jma}.  We compute the $ggH$ and $ggHH$ amplitudes in the
limit where both the effective and the full theory are valid, i.e.  for small
external momenta as compared to the top quark mass. This leads again to
single-scale vacuum integrals up to four-loop order. In the following
we use $\alpha_s \equiv \alpha_s^{(5)}(\mu)$ if not indicated differently.

\begin{table}[t]
  \begin{center}
    \begin{tabular}[t]{c||r|r|r|r}
      \# loops & 1 & 2 & 3 & 4 \\
      \hline
      $ggH$        &   2 &   23 &   657&   23\,251 \\
      $ggHH$ 1PI   &   6 &   99 &3\,192&  124\,149 \\
      $ggHH$ 1PR   & --- &    8 &  216 &    7\,200 \\
    \end{tabular}
    \caption{\label{tbl::numdia_ggh_gghh}Number of diagrams needed for 
      computing the Higgs-gluon amplitudes up to four loops.}
  \end{center}
\end{table}

%- {{{ C_H:

\subsection{$C_H$}

The Wilson coefficient is obtained by comparing the $ggH$ amplitude in the
effective and full theory which leads to the following matching formula
\begin{align}
  C_H Z_{\mathcal{O}_1} \mathcal{A}^\mathrm{eff}_\mathrm{LO} =
  \frac{1}{\zeta_3^0}\mathcal{A}^h + {\cal O}(1/m_t)\,.
  \label{eq::match_CH}
\end{align}
On the full-theory side $\mathcal{A}^h$ denotes the hard part of the
amplitude, which is obtained from a Taylor expansion in the two external
momenta.  It is assumed that the top quark mass and $\alpha_s$ are renormalized
using standard counterterms up to three loops and the factor $1/\zeta_3^0$
takes care of the non-vanishing part of the gluon wave function
renormalization.  Due to our choice of the kinematic variables there are only
tree-level contributions on the effective-theory side. Furthermore, we have
the renormalization constant of the effective operator, $Z_{\mathcal{O}_1}$,
and the sought-after (renormalized) matching coefficient $C_H$, which is
obtained by dividing Eq.~(\ref{eq::match_CH}) by $Z_{\mathcal{O}_1}$.  Note
that $Z_{\mathcal{O}_1}$ depends on $\alpha_s^{(5)}$ whereas the quantities on
the r.h.s. depend on $\alpha_s^{(6)}$. Before combining the various parts we
use the decoupling constant to transform the strong coupling constant to
$\alpha_s^{(5)}$.  We renormalize the top quark mass in a first step in the
$\overline{\rm MS}$ scheme and transform to the on-shell scheme afterwards.

\begin{figure}[t]
  \begin{center}
    \includegraphics[width=0.2\textwidth]{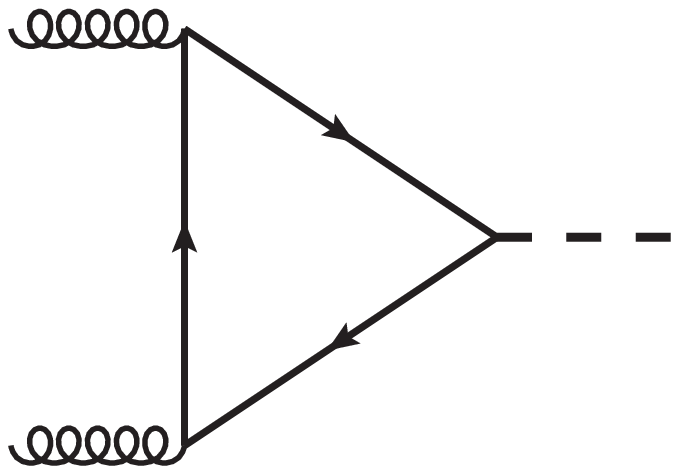}
    \includegraphics[width=0.2\textwidth]{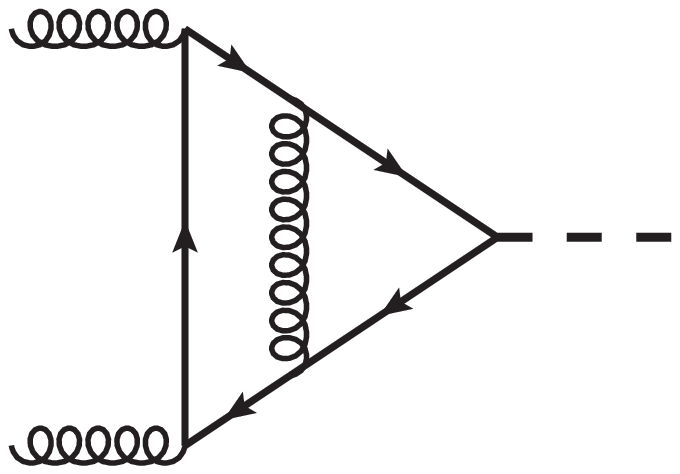}
    \includegraphics[width=0.2\textwidth]{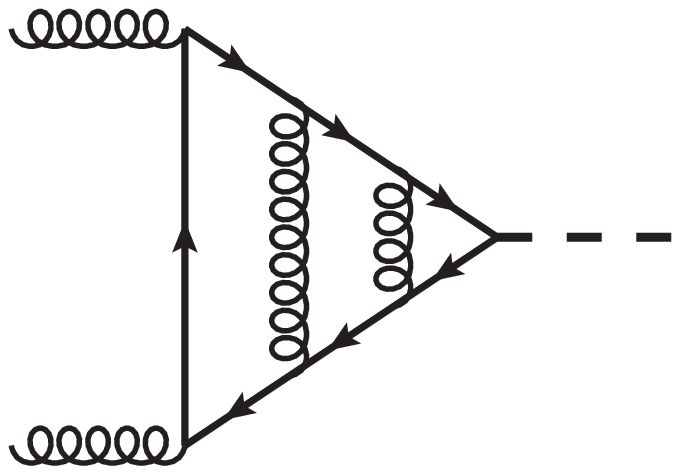}
    \includegraphics[width=0.2\textwidth]{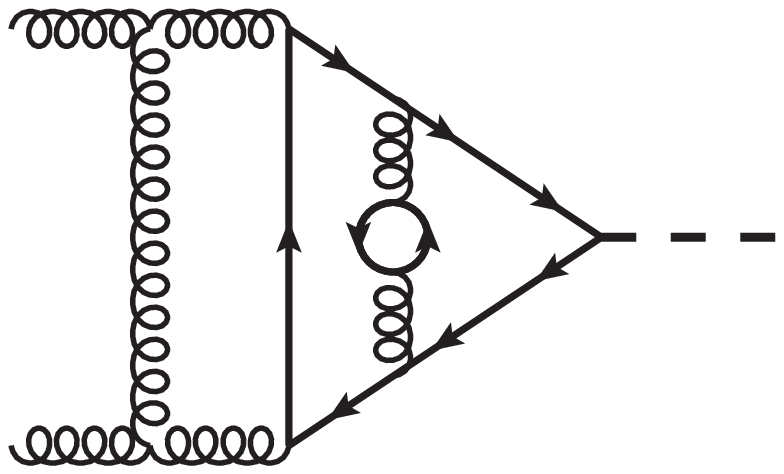}
    \\
    \includegraphics[width=0.2\textwidth]{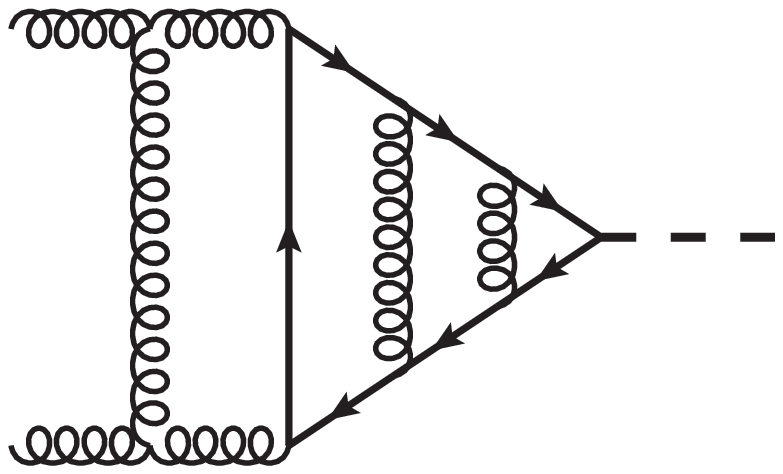}
    \includegraphics[width=0.2\textwidth]{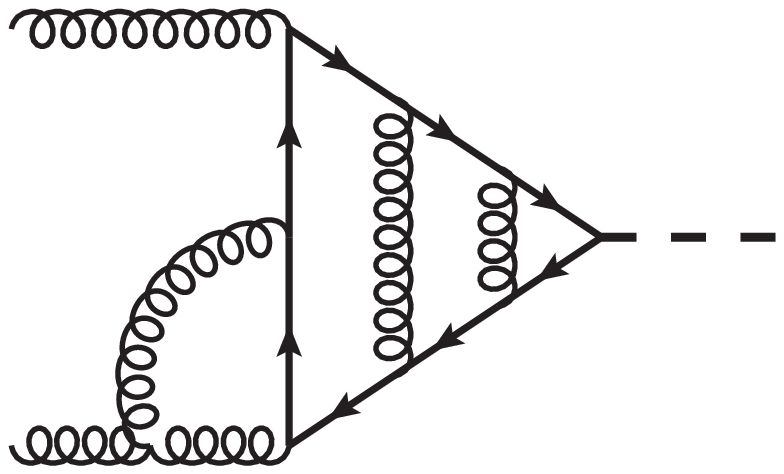}
    \includegraphics[width=0.2\textwidth]{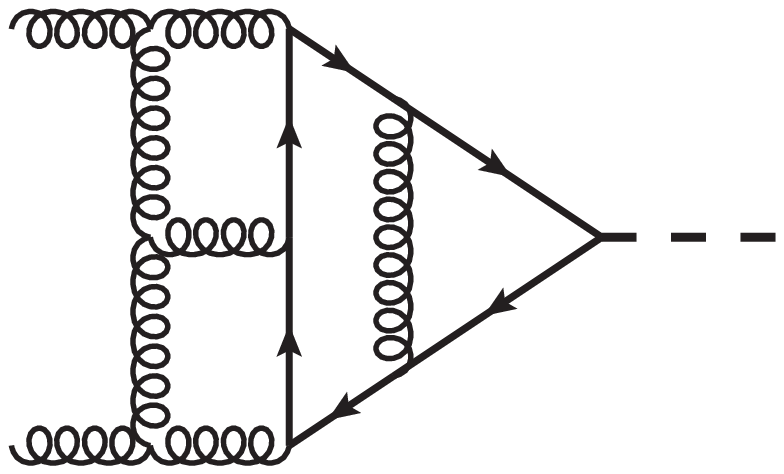}
    \includegraphics[width=0.2\textwidth]{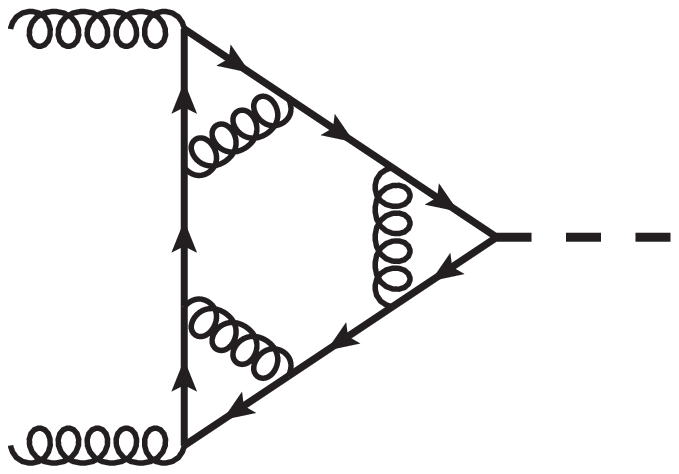}
  \end{center}
\caption{\label{fig::gghfull} Sample one-, two-, three- and four-loop diagrams
  contributing to the $gg \rightarrow H$ amplitude. Solid and curly lines refer to
    fermions and gluons, respectively. The external Higgs boson is represented
    by a dashed line.}
\end{figure}

The number of diagrams generated by \verb|qgraf| for $\mathcal{A}^h$ is shown
in Tab.~\ref{tbl::numdia_ggh_gghh} and sample Feynman diagrams
are shown in Fig.~\ref{fig::gghfull}. In a first step we apply
the projector
\begin{align}
  P^{\mu\nu} = \frac{1}{2-2\epsilon}\left(g^{\mu\nu}q_1\cdot q_2 
  - q_1^\nu q_2^\mu - q_1^\mu q_2^\nu \right)\,,
  \label{eq::projch}
\end{align} where $q_1^\mu$ and $q_2^\nu$ are the incoming four-momenta of the
external gluons with polarization vectors $\varepsilon^\mu(q_1)$ and
$\varepsilon^\nu(q_2)$. After tensor reduction we obtain the same kind of integral
families as for the decoupling constants of the previous section.

As before, we perform the calculation for generic SU($N_c$) colour factors 
and full dependence on the gauge parameter $\xi$, which drops out after
summing all contributions to $\mathcal{A}^h$.
We cast the final result for the Wilson coefficient $C_H$ in the form
\begin{align}
  C_H = -\frac{2\alpha_s}{3\pi} T_F \sum_{i = 1} C_H^{(i)}\left(\frac{\alpha_s}{\pi}\right)^{(i-1)}\,,
\end{align}
where the $C^{(i)}$ are given by
\begin{align}
  C_H^{(1)} &= 1~,\\
  C_H^{(2)} &= \frac{5}{4} C_A - \frac{3}{4} C_F~,\\
  C_H^{(3)} &= \frac{1063}{576} C_A^2 - \frac{25}{12} C_A C_F - \frac{5}{96}
              C_A T_F + \frac{27}{32} C_F^2 - \frac{1}{12} C_F T_F\nonumber\\
            &+ \left[\frac{7}{16} C_A^2 - \frac{11}{16} C_A
              C_F\right]\ln\left(\frac{\mu^2}{M_t^2}\right)  + n_l
              T_F\left[-\frac{47}{144} C_A - \frac{5}{16} C_F
              + \frac{1}{2} C_F\ln\left(\frac{\mu^2}{M_t^2}\right)\right]~,\\
  C_H^{(4)} &= C_A^3\left(\frac{110041}{41472} - \frac{1577}{3072} \zeta(3)\right) + C_A^2 C_F\left(- \frac{{ 99715}}{6912} + \frac{5105}{512}\zeta(3)\right)\nonumber\\
            &+ C_A^2 T_F\left(- \frac{1081}{3456} + \frac{1}{384}\zeta(3)\right) + C_A C_F^2\left(\frac{{ 2963}}{384} - \frac{407}{128}\zeta(3)\right)\nonumber\\ 
            &+ C_A C_F T_F\left(\frac{4537}{1728} - \frac{115}{64}\zeta(3)\right) + C_A T_F^2\left(\frac{2}{27} - \frac{7}{64}\zeta(3)\right) - \frac{471}{128} C_F^3\nonumber\\
            &+ C_F^2 T_F\left(- \frac{5}{12} + \frac{13}{32}\zeta(3)\right) + C_F T_F^2\left(\frac{113}{432} - \frac{7}{32}\zeta(3)\right)\nonumber\\
            &+ \frac{d_R^{abcd}d_A^{abcd}}{N_A { T_F}}\left(- { \frac{2}{3}} +
              { \frac{13}{2}}\zeta(3)\right) + \frac{d_R^{abcd}d_R^{abcd}}{N_A { T_F}}\left(\frac{11}{{ 12}} - { 2}\zeta(3)\right)\nonumber\\
            &+ \left[\frac{1993}{1152} C_A^3 - { \frac{275}{72}} C_A^2 C_F - \frac{55}{576} C_A^2 T_F + \frac{99}{{ 64}} C_A C_F^2 - \frac{11}{72} C_A C_F T_F\right]\ln\left(\frac{\mu^2}{M_t^2}\right)\nonumber\\
            &+ \left[\frac{77}{192} C_A^3 - \frac{121}{192} C_A^2
              C_F\right]\ln^2\left(\frac{\mu^2}{M_t^2}\right)+ n_l\frac{d_R^{abcd}d_R^{abcd}}{N_A { T_F}}\left(\frac{11}{{ 6}} - { 4}\zeta(3)\right)\nonumber\\
            &+ n_l T_F\Bigg[C_A^2\left(- \frac{12421}{10368} - \frac{151}{256}\zeta(3)\right) + C_A C_F\left(\frac{9605}{2592} - \frac{1145}{384}\zeta(3)\right)\nonumber\\
            &+ C_A T_F\left(\frac{7}{216} - \frac{7}{64}\zeta(3)\right)+ C_F^2\left(\frac{{ 215}}{288} + \frac{127}{96}\zeta(3)\right)+ C_F T_F\left(- \frac{29}{144} - \frac{7}{32}\zeta(3)\right)\Bigg] 
           \nonumber\\
            &+ n_l^2 T_F^2\left[-\frac{161}{2592} C_A - \frac{677}{1296} C_F\right]
              \nonumber\\
            &+ n_l T_F\left[- \frac{55}{288} C_A^2 + \frac{55}{36} C_A C_F + \frac{5}{144} C_A T_F - { \frac{5}{8}} C_F^2 + \frac{1}{18} C_F T_F\right]\ln\left(\frac{\mu^2}{M_t^2}\right)\nonumber\\
            &+ n_l^2 T_F^2\left[\frac{5}{144} C_A + \frac{1}{18} C_F\right]\ln\left(\frac{\mu^2}{M_t^2}\right) 
              + n_l T_F\left[- \frac{7}{48} C_A^2 + \frac{11}{16} C_A C_F\right]\ln^2\left(\frac{\mu^2}{M_t^2}\right)\nonumber\\
            &- \frac{1}{6} n_l^2 C_F T_F^2\ln^2\left(\frac{\mu^2}{M_t^2}\right)\,.
\label{eq::CH}
\end{align}
$\zeta(n)$ is the Riemann $\zeta$-function, evaluated at $n$, $M_t$ is the
on-shell top quark mass and the  SU$(N_c)$ colour factors are given by
\begin{align}
  &C_A = N_c,\qquad C_F = \frac{N_c^2-1}{2N_c},\qquad T_F = \frac{1}{2},\nonumber\\
  &\frac{d_R^{abcd}d_A^{abcd}}{N_A} = \frac{N_c(N_c^2+6)}{48},\qquad\frac{d_R^{abcd}d_R^{abcd}}{N_A} = \frac{N_c^4 - 6 N_c^2 + 18}{96 N_c^2}\,,
    \label{eq::cf}
\end{align}
with $N_A=N_c^2-1$. Note that $C_H$ only contains
$\zeta(3)$ as a transcendental constant while $\mathcal{A}^h$ also contains
other zeta-values and polylogarithms up to weight four.  They cancel
in the combination with $1/\zeta_3^0$ and only $\zeta(3)$ survives.  After
specifying $N_c=3$ our
result is in full agreement with the expression obtained with the help of the
LET~\cite{Chetyrkin:1997un}. The latter can be used to obtain the five-loop
result with full colour structure. We refrain from showing explicit results in
the paper but include them in the ancillary files~\cite{progdata}.  Let
us remark that the five-loop result contains zeta-values and polylogarithms up
to weight five.

%- }}}
%- {{{ C_HH:

\subsection{$C_{HH}$}

\begin{figure}[t]
  \begin{center}
    \includegraphics[width=.27\textwidth]{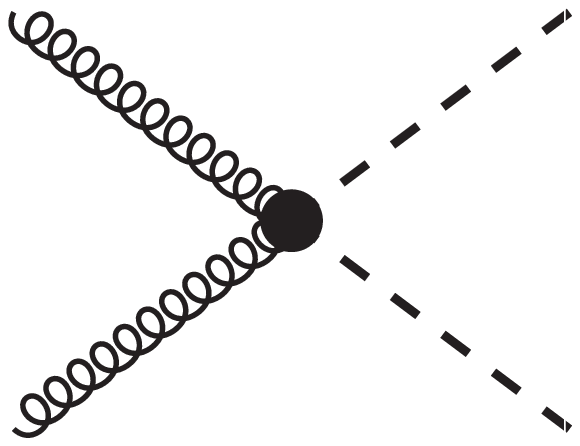}
    \includegraphics[width=.25\textwidth]{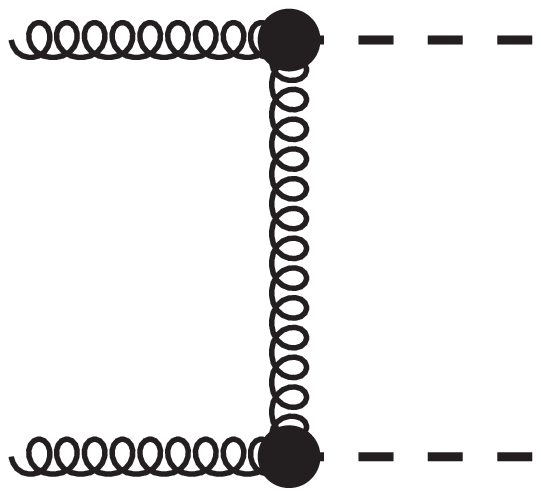}
    \includegraphics[width=.4\textwidth]{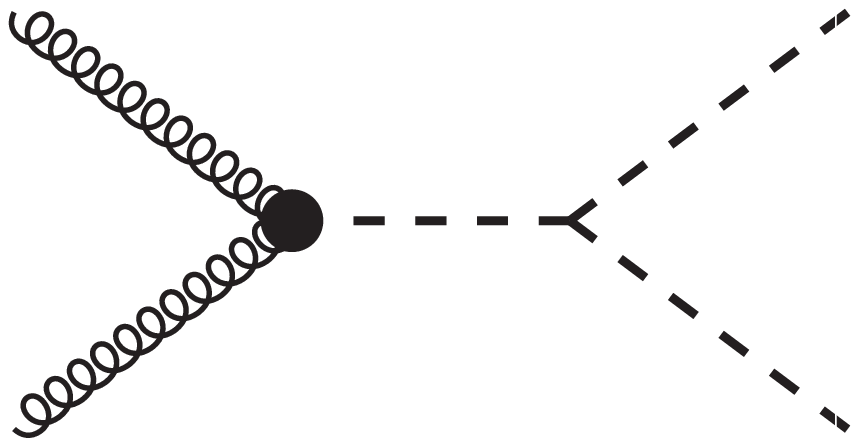}
    \caption{\label{fig::gghheff}Tree-level contributions to the
      $gg \rightarrow HH$ amplitude in the effective theory.  The blob
      indicates the insertion of the operator ${\cal O}_1$. The left diagram
      is proportional to $C_{HH}$, the one in the middle to $C_H^2$ and the
      right diagram, which contains the trilinear Higgs coupling $\lambda$, to
      $C_H$.  The amplitudes corresponding to the three Feynman diagrams are
      denoted by $\mathcal{A}^\mathrm{eff}_{\mathrm{LO,1PI}}$,
      $\mathcal{A}^\mathrm{eff}_{\mathrm{LO,1PR,}\lambda = 0}$ and
      $\mathcal{A}^\mathrm{eff}_{\mathrm{LO,1PR,}\lambda \neq 0}$.}
  \end{center}
\end{figure}

The matching procedure to obtain $C_{HH}$ is more involved as for $C_H$.
First of all there are three contributions on the effective-theory side which
are shown in Fig.~\ref{fig::gghheff}: a one-particle irreducible (1PI) term
proportional to $C_{HH}$, a one-particle reducible (1PR) term, which involves
$C_H^2$, and a term mediated by a virtual Higgs boson which splits into a
Higgs boson pair via the Higgs boson self-coupling $\lambda$. The latter is
similar in nature to the effective amplitude in the matching formula for
$C_H$.  In fact, also on the full-theory side this contribution involves
diagrams which we already encountered in the computation of $C_H$. As
mentioned in Ref.~\cite{Grigo:2014jma} it is easy to see, that these diagrams
exactly cancel between the full and effective theory.
Thus, the contributions relevant to extract $C_{HH}$ are the 1PI
and 1PR contribution with $\lambda=0$.

The effective-theory side of the matching formula is obtained after
renormalizing the operators in the various contributions of
Fig.~\ref{fig::gghheff}. Whereas the left and right contributions are both
renormalized with a factor $Z_{\mathcal{O}_1}$, the term in the middle needs
special care. In fact, a naive renormalization with $(Z_{\mathcal{O}_1})^2$
leads to uncanceled poles as has already been observed in
Refs.~\cite{Zoller:2012qv,Zoller:2014dca}. A careful analysis of the
renormalization of the product of two operators $\mathcal{O}_1$ has been
performed in Ref.~\cite{Zoller:2016iam} along the lines
of~\cite{Spiridonov:1984br}. It has been observed that apart from the naive
multiplicative renormalization a further term is needed which is proportional to
a single $\mathcal{O}_1$. Adapting the findings of Ref.~\cite{Zoller:2016iam}
to our notation one has
\begin{align}
  \mathcal{A}^\mathrm{eff}_{(\mathcal{O}_1)^2} = Z_{\mathcal{O}_1}^2
  \mathcal{A}^\mathrm{eff}_{(\mathcal{O}_1^0)^2} +
  Z_{11}^L\mathcal{A}^\mathrm{eff}_{\mathcal{O}_1^0}~\,
  \label{eq::renO12}
\end{align}
where $\mathcal{A}^\mathrm{eff}_{\mathcal{O}_1^0}$ and
$\mathcal{A}^\mathrm{eff}_{(\mathcal{O}_1^0)^2}$ correspond to amplitudes with
one and two operator insertions.  The renormalization constant $Z_{11}^L$
(where $L$ stands for ``linear'') is given by~\cite{Zoller:2016iam}
\begin{align}
  Z_{11}^L = \frac{1}{\epsilon}\left(1 -
  \frac{\beta(\alpha_s)}{\epsilon}\right)^{-2}\alpha_s^2
  \frac{\partial}{\partial\alpha_s}\left[\frac{\beta(\alpha_s)}{\alpha_s}\right]\,. 
\end{align}
It has its first non-vanishing contribution at order $\alpha_s^2$.
As we will see below, in our calculation we need the combination
$Z_{11}^L/Z_{\mathcal{O}_1}$ up
to order $\alpha_s^2$ which is given by
\begin{align}
  \frac{Z_{11}^L}{Z_{\mathcal{O}_1}} = -\frac{\beta_1}{\epsilon}\frac{\alpha_s^2}{(4\pi)^2} + \mathcal{O}(\alpha_s^3)\,.
\end{align}

\begin{figure}[t]
  \begin{center}
    \includegraphics[width=0.2\textwidth]{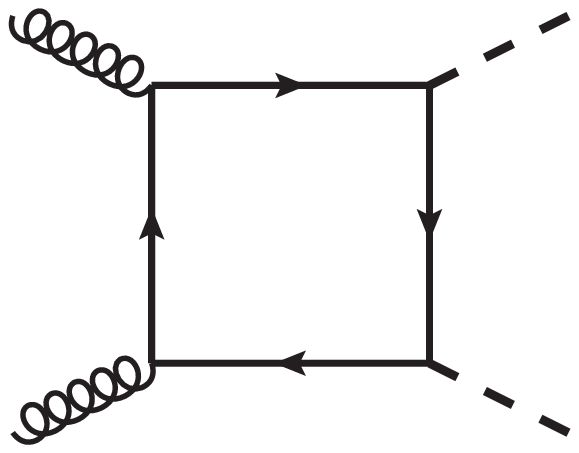}
    \includegraphics[width=0.2\textwidth]{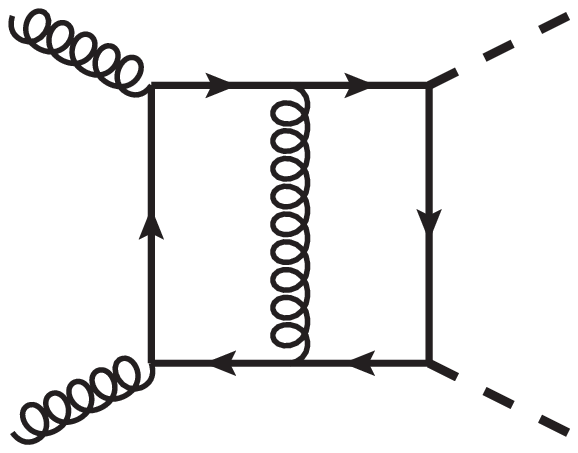}
    \includegraphics[width=0.2\textwidth]{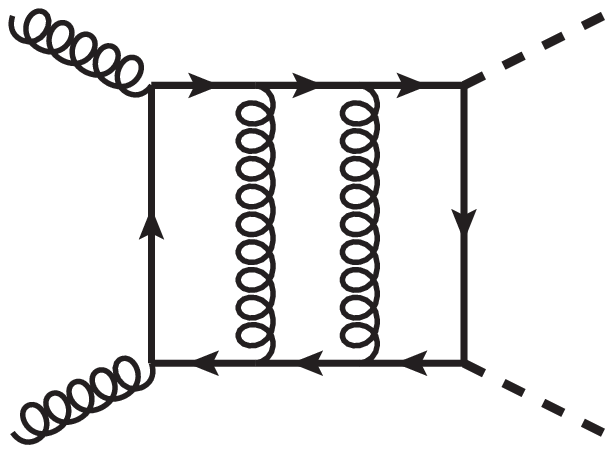}
    \\
    \includegraphics[width=0.2\textwidth]{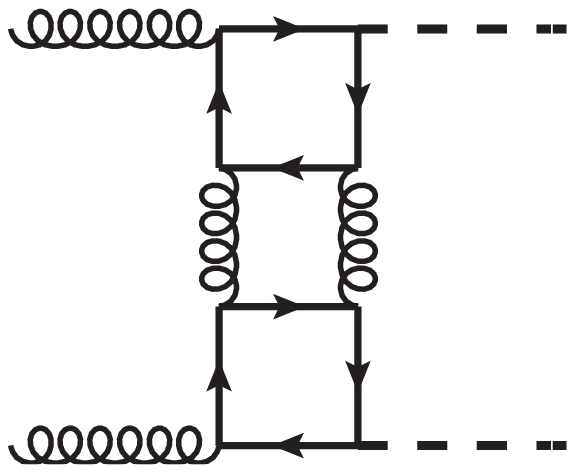}
    \includegraphics[width=0.2\textwidth]{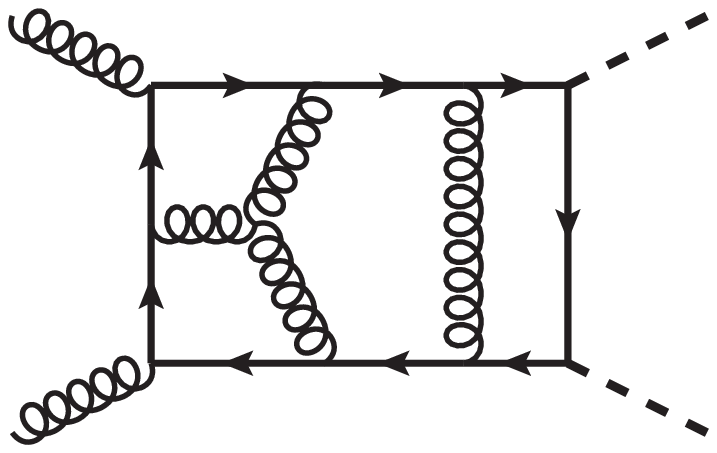}
    \includegraphics[width=0.2\textwidth]{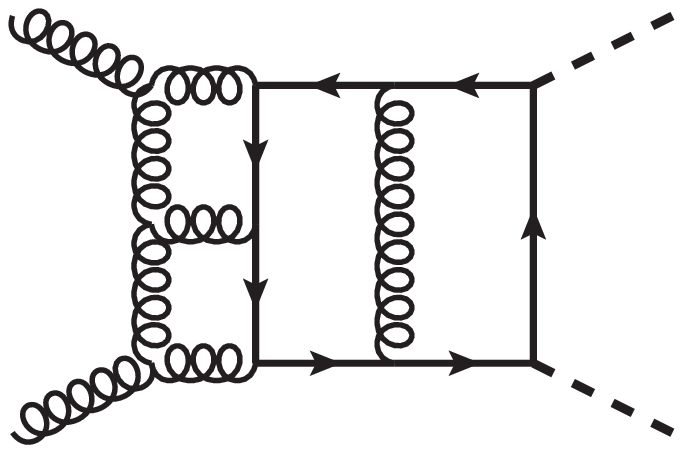}
  \end{center}
  \caption{\label{fig::gghh1PIfull} Sample one-, two-, three- and four-loop
    diagrams contributing $\mathcal{A}^h_{\mathrm{1PI}}$ in
    Eq.~(\ref{eq::match_CHH}).}
\end{figure}

\begin{figure}[t]
  \begin{center}
    \includegraphics[width=0.2\textwidth]{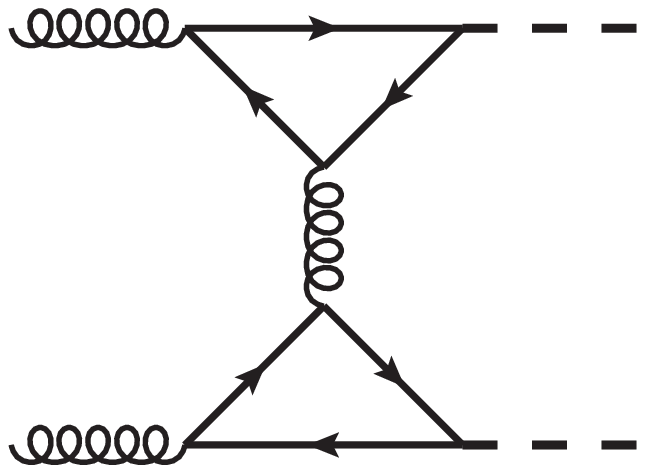}
    \includegraphics[width=0.2\textwidth]{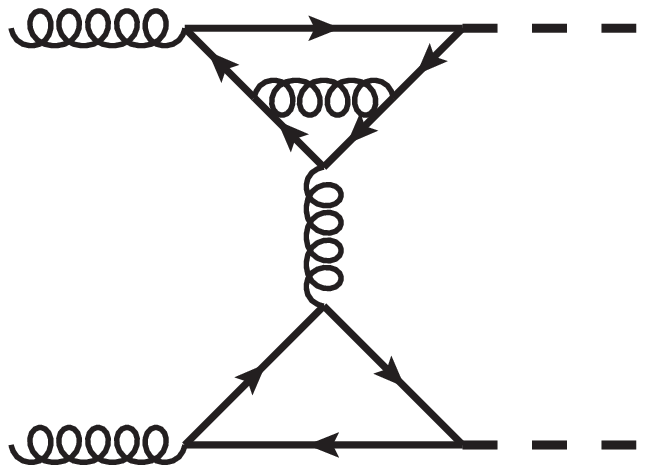}
    \includegraphics[width=0.2\textwidth]{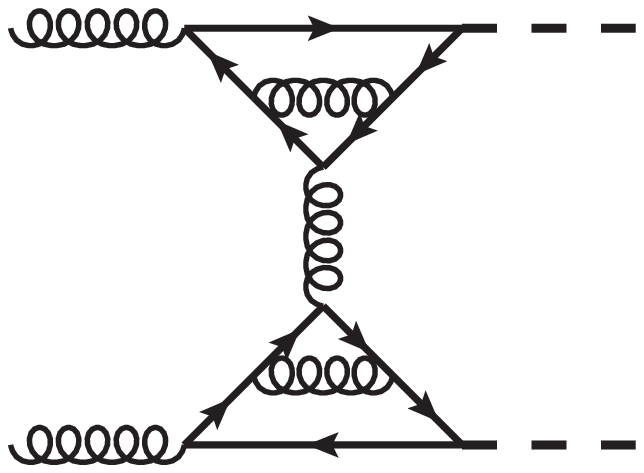}
    \includegraphics[width=0.2\textwidth]{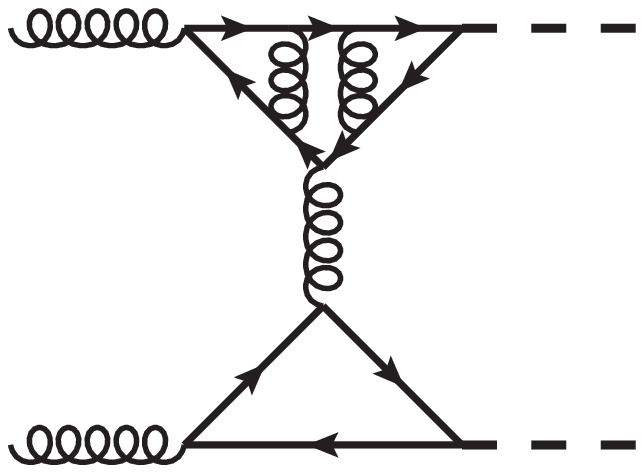}
  \end{center}
  \caption{\label{fig::gghh1PRfull} Sample two-, three- and four-loop diagrams
    contributing to $\mathcal{A}^h_{\mathrm{1PR,}\lambda = 0}$ in
    Eq.~(\ref{eq::match_CHH}).}
\end{figure}

We are now in the position to write down the matching formula for $C_{HH}$.
Complementing the effective-theory side, which is basically given by
Fig.~\ref{fig::gghheff}, with the corresponding full-theory amplitudes
and taking into account Eq.~(\ref{eq::renO12}) leads to\footnote{When
  applying Eq.~(\ref{eq::renO12}) to Higgs boson pair production we have
  $\mathcal{A}^\mathrm{eff}_{(\mathcal{O}_1^0)^2} =
  \mathcal{A}^\mathrm{eff}_{\mathrm{LO,1PR,}\lambda = 0}$
  and $\mathcal{A}^\mathrm{eff}_{\mathcal{O}_1^0} =
  \mathcal{A}^\mathrm{eff}_{\mathrm{LO,1PI}}$.}
\begin{align}
  &(C_{HH}Z_{\mathcal{O}_1} +
    C_H^2Z_{11}^L)\mathcal{A}^\mathrm{eff}_{\mathrm{LO,1PI}} +
    C_H^2Z_{\mathcal{O}_1}^2
    \mathcal{A}^\mathrm{eff}_{\mathrm{LO,1PR,}\lambda = 0} +
    C_HZ_{\mathcal{O}_1}
    \mathcal{A}^\mathrm{eff}_{\mathrm{LO,1PR,}\lambda \neq
    0}\nonumber\\ 
  &= \frac{1}{\zeta_3^0}\left(\mathcal{A}^h_{\mathrm{1PI}} +
    \mathcal{A}^h_{\mathrm{1PR,}\lambda = 0} +
    \mathcal{A}^h_{\mathrm{1PR,}\lambda \neq 0}\right) 
    + {\cal O}(1/m_t)
    \,,
    \label{eq::match_CHH}
\end{align}
where sample Feynman diagrams contributing to $\mathcal{A}^h_{\mathrm{1PI}}$
and $\mathcal{A}^h_{\mathrm{1PR,}\lambda = 0}$ can be found in Figs.~\ref{fig::gghh1PIfull}
and~\ref{fig::gghh1PRfull}, respectively. As already mentioned above, the 
contributions with $\lambda\neq0$ cancel in Eq.~(\ref{eq::match_CHH}).
Note that our matching formula differs from the one of Ref.~\cite{Grigo:2014jma}
by the term proportional to $Z_{11}^L$ which contributes for the first time
at four-loop order, since both $C_H^2$ and $Z_{11}^L$ are of order $\alpha_s^2$.

Let us in the following discuss some features of the matching procedure.
At one-loop order the only non-zero contribution on the r.h.s.
of Eq.~(\ref{eq::match_CHH}) is $\mathcal{A}^h_{\mathrm{1PI}}$ and one obtains
$C_{HH}^{(1)} = C_H^{(1)}$. This also holds at two-loops where the 1PR contributions
on effective- and full-theory side match exactly.
A non-trivial interplay between $\mathcal{A}^h_{\mathrm{1PI}}$ and
$\mathcal{A}^h_{\mathrm{1PR,}\lambda = 0}$ is observed for the first time at
three-loop order~\cite{Grigo:2014jma}. In fact the 1PI and 1PR contributions
are not separately finite any more and the poles only cancel in the sum.
Starting from this order $C_{HH}$ is different from $C_H$.
While the 1PI and 1PR contributions are separately $\xi$-independent at three loops,
for the four-loop colour structure $C_A^2 T_F$  it only drops out in the proper combination.

We computed the 1PI and 1PR amplitudes in the full theory separately and keep
in both cases terms linear in the gauge parameter $\xi$.  For both
contributions it is important to keep the three external momenta different
from zero and different from each other in order to avoid the mixing with
unphysical operators~\cite{Spiridonov:1984br}. The external momenta
can be set to zero after projection to the matching
coefficient which is done with the help of
\begin{align}
  P^{\mu\nu} = &\frac{1}{2-4\epsilon}\left(\frac{q_1^\nu q_2^\mu
                 q_{33}}{2q_{12}q_T^2} - \frac{q_1^\nu q_2^\mu}{2q_{12}} -
                 \frac{q_1^\nu q_3^\mu q_{23}}{q_{12}q_T^2} 
                 -\frac{q_2^\mu q_3^\nu q_{13}}{q_{12}q_T^2} + \frac{q_3^\mu
                 q_3^\nu}{q_T^2} + g^{\mu\nu}\right)\nonumber\\ 
  &-\frac{q_1^\nu q_2^\mu q_{33}}{4q_{12}q_T^2} - \frac{q_1^\nu
     q_2^\mu}{4q_{12}} + \frac{q_1^\nu q_3^\mu q_{23}}{2q_{12}q_T^2} 
     +\frac{q_2^\mu q_3^\nu q_{13}}{2q_{12}q_T^2} - \frac{q_3^\mu q_3^\nu}{2q_T^2}~,
    \label{eq::proj_chh}
\end{align}
where $q_{ij} = q_i \cdot q_j$ and $q_T^2 = 2q_{13}q_{23}/q_{12} - q_{33}$.
$q_1^\mu$ and $q_2^\nu$ are the incoming four-momenta of the external gluons
with polarization vectors $\varepsilon^\mu(q_1)$ and $\varepsilon^\nu(q_2)$
and $q_3$ is the incoming four-momentum of one of the Higgs bosons.

The number of diagrams for the 1PI amplitude can be found in
Tab.~\ref{tbl::numdia_ggh_gghh} and sample diagrams are shown in
Fig.~\ref{fig::gghh1PIfull}. Once the projector of Eq.~(\ref{eq::proj_chh}) is
applied one obtains scalar expressions which still contain scalar products of
$q_1$, $q_2$ and $q_3$ and loop momenta in the numerator.  After solving the
corresponding tensor vacuum integrals the resulting scalar products $q_{ij}$
cancel against the corresponding contributions with negative powers from the
projector and all external momenta can be set to zero.

The 1PR amplitude has been obtained in two different ways. First, we computed
the 1PR diagrams up to four-loop order (see Tab.~\ref{tbl::numdia_ggh_gghh}
for the number of diagrams and Fig.~\ref{fig::gghh1PRfull} for typical Feynman
diagrams) in analogy to the 1PI contribution.  As a cross-check we computed
the 1PI parts of the 1PR contributions separately and constructed the $n$-loop
1PR $ggHH$ amplitude from $ggH$ amplitudes computed up to $(n-1)$ loops. In this
approach one of the gluons in the $ggH$ amplitude has to be off-shell, which
leads to more non-vanishing Lorentz structures. In practice, we computed the
1PI $ggH$ amplitudes with open Lorentz indices up to three loops.
Full agreement has been found between the two methods.

We cast the final result for the Wilson coefficient $C_{HH}$ in the form
\begin{align}
C_{HH} = -\frac{2\alpha_s}{3\pi} T_F \sum_{i = 1} \left(C_{H}^{(i)} + \Delta_{HH}^{(i)}\right)\left(\frac{\alpha_s}{\pi}\right)^{(i-1)}~,
\end{align}
where the $C_H^{(i)}$ are given in Eq.~\eqref{eq::CH} and
the differences are given by
\begin{align}
  \Delta_{HH}^{(1)} &= 0~,\nonumber\\
  \Delta_{HH}^{(2)} &= 0~,\nonumber\\
  \Delta_{HH}^{(3)} &= \frac{7}{8} C_A^2 - \frac{11}{8} C_A C_F - \frac{5}{6}
                      C_A T_F + \frac{1}{2} C_F T_F + n_l C_F T_F
                      ~,\nonumber\\ 
  \Delta_{HH}^{(4)} &= \frac{1993}{576} C_A^3 - \frac{1289}{144} C_A^2 C_F -
                      \frac{3191}{864} C_A^2 T_F + \frac{165}{32} C_A C_F^2 +
                      \frac{67}{18} C_A C_F T_F + \frac{5}{72} C_A
                      T_F^2\nonumber\\ &- \frac{3}{2} C_F^2 T_F + \frac{1}{9}
                                         C_F T_F^2 + \left[\frac{77}{48} C_A^3
                                         - \frac{121}{48} C_A^2 C_F -
                                         \frac{7}{12} C_A^2 T_F +
                                         \frac{11}{12} C_A C_F
                                         T_F\right]\ln\left(\frac{\mu^2}{M_t^2}\right)\nonumber\\  
                    & + n_l T_F \left[- \frac{55}{144} C_A^2 + \frac{55}{18} C_A C_F +
                      \frac{109}{216}C_A T_F -
                      \frac{11}{4} C_F^2
                      +  \frac{19}{36} C_F
                      T_F\right]\nonumber\\  
                    & + n_l^2 T_F^2 \left[\frac{5}{72} C_A + \frac{1}{9}
                      C_F\right] + n_l T_F \left[- \frac{7}{12} C_A^2 +
                      \frac{11}{4} C_A C_F - \frac{2}{3} C_F
                      T_F\right]\ln\left(\frac{\mu^2}{M_t^2}\right)\nonumber\\  
                    & - \frac{2}{3} n_l^2 C_F T_F^2 \ln\left(\frac{\mu^2}{M_t^2}\right)\,.
\end{align}
The three-loop result can be found in Ref.~\cite{Grigo:2014jma}.  Our
four-loop result $\Delta_{HH}^{(4)}$ agrees with the expression from
Eq.~(\ref{eq::let_chh})~\cite{Spira:2016zna}. We can thus confirm the validity
of the LET for $C_{HH}$~\cite{Spira:2016zna} through four loops.  In analogy
to $C_H$ also for $C_{HH}$ it is possible to construct the five-loop
approximation for general colour structure. The corresponding results can be
found in computer readable form in the ancillary files~\cite{progdata}.  After
specifying to $N_c=3$ we agree with the numerical results given in
Ref~\cite{Spira:2016zna}, both for $\overline{\rm MS}$ and on-shell top quark
mass.

%- }}}

%- }}}
%- {{{ Conclusions:

\section{\label{sec:conclusions}Conclusions}

We perform for the first time a direct four-loop computation of the Wilson
coefficients $C_H$ and $C_{HH}$ of the effective operators, which couple
gluons to one and two Higgs bosons, respectively. $C_H$ and $C_{HH}$ enter as
building blocks various physical quantities, e.g., the
next-to-next-to-next-to-leading order predictions for
single~\cite{Anastasiou:2016cez,Mistlberger:2018etf} and double Higgs boson
production.\footnote{See also the recent paper~\cite{Banerjee:2018} where
  two-loop massless four-point amplitudes have been computed, a further
  building block to next-to-next-to-next-to-leading order double Higgs boson
  production.} Our results for $C_H$ and $C_{HH}$ agree with the expression
obtained by means of LETs. Furthermore, we compute all QCD decoupling
constants up to four-loop order. If possible we compared with the literature
and find agreement after specifying the colour factors. All our results are
expressed for general SU$(N_c)$ colour factors whereas the four-loop
expressions in the literature are only available for $N_c=3$.

A major result of this paper is the derivation of the matching
equation~(\ref{eq::match_CHH}) which receives a non-trivial renormalization
contribution from the effective-theory amplitude with two insertions
of the operator ${\cal O}_1$. The new term contributes for the first time
at four-loop order and is essential to obtain a finite result.

For the convenience of the reader we collect all analytic results
obtained in this paper in ancillary files~\cite{progdata}.

%- }}}
%- {{{ Acknowledgements:

\section*{Acknowledgements}

We thank Kirill Melnikov and Konstantin Chetyrkin for fruitful discussions and
Konstantin Chetyrkin for drawing our attention to Ref.~\cite{Zoller:2016iam}.
F.H. acknowledges the support by the DFG-funded Doctoral School KSETA.

%- }}}
%- {{{ Appendix:

\begin{appendix}

\section{\label{app:dec}Decoupling constants}

In this appendix we collect the results for the decoupling constants
for general SU($N_c$) colour factors. We provide results for
$\zeta_{\alpha_s}$ and $\zeta_m$ up to four loops and show for all
other $\zeta$s the expressions for the first non-vanishing loop-order.
Computer readable expressions up to four loops can be found in~\cite{progdata}.
Our results read
\begin{align}
\zeta_X = 1 + \sum_{i=1}\zeta_X^{(i)}\left(\frac{\alpha^{(n_f)}_s}{\pi}\right)^{i}~,
\end{align}
where
\begin{align}
\zeta_{\alpha_s}^{(1)} &= - \frac{1}{3} T_F \ln\left(\frac{\mu^2}{m^2}\right)~,\nonumber\\
\zeta_{\alpha_s}^{(2)} &= \frac{2}{9} C_A T_F - \frac{13}{48} C_F T_F + \Big(- \frac{5}{12} C_A T_F + \frac{1}{4} C_F T_F\Big)\ln\left(\frac{\mu^2}{m^2}\right) + \frac{1}{9} T_F^2\ln^2\left(\frac{\mu^2}{m^2}\right)~,\nonumber\\
\zeta_{\alpha_s}^{(3)} &= C_A^2 T_F\Big(\frac{11347}{20736} - \frac{5}{1536}\zeta(3)\Big) + C_A C_F T_F\Big(\frac{2999}{2592} - \frac{1273}{768}\zeta(3)\Big) + C_A T_F^2\Big(\frac{245}{5184}\nonumber\\ & - \frac{7}{128}\zeta(3)\Big) + C_F^2 T_F\Big(- \frac{97}{288} + \frac{95}{192}\zeta(3)\Big) + C_F T_F^2\Big(\frac{103}{1296} - \frac{7}{64}\zeta(3)\Big)\nonumber\\ & + n_l T_F\Bigg[-\frac{1}{2592} C_A T_F - \frac{41}{162}C_F T_F\Bigg] + \Bigg[- \frac{1063}{1728} C_A^2 T_F + \frac{25}{36} C_A C_F T_F - \frac{113}{864} C_A T_F^2\nonumber\\ & - \frac{9}{32} C_F^2 T_F + \frac{5}{24} C_F T_F^2\Bigg]\ln\left(\frac{\mu^2}{m^2}\right)+\Bigg[- \frac{7}{96} C_A^2 T_F + \frac{11}{96} C_A C_F T_F + \frac{25}{72} C_A T_F^2\nonumber\\ & - \frac{5}{24} C_F T_F^2\Bigg]\ln^2\left(\frac{\mu^2}{m^2}\right) - \frac{1}{27} T_F^3\ln^3\left(\frac{\mu^2}{m^2}\right) + n_l T_F\Bigg[\frac{47}{432} C_A T_F + \frac{5}{48} C_F T_F\Bigg]\ln\left(\frac{\mu^2}{m^2}\right)\nonumber\\ & - \frac{1}{12} n_l C_F T_F^2  \ln^2\left(\frac{\mu^2}{m^2}\right)~,\nonumber\\
\zeta_{\alpha_s}^{(4)} &= C_A^3 T_F\Big(
          \frac{14060183}{13063680}
          - \frac{4663}{630} \mathrm{Li}_5\left(\frac{1}{2}\right)
          + \frac{24153}{2240} \mathrm{Li}_4\left(\frac{1}{2}\right)
          + \frac{8051}{17920} \ln^4(2)\nonumber\\
          &+ \frac{4663}{75600} \ln^5(2)
          + \frac{377777}{40320} \zeta(5)
          - \frac{6668653}{645120} \zeta(4)
          - \frac{70841}{10080} \zeta(4) \ln(2)
          + \frac{1331653}{215040} \zeta(3)\nonumber\\
          &- \frac{24153}{8960} \zeta(2) \ln^2(2)
          - \frac{4663}{7560} \zeta(2) \ln^3(2)
          \Big)+ C_A^2 C_F T_F \Big(
          \frac{69024559}{10450944} 
          + \frac{8674}{315} \mathrm{Li}_5\left(\frac{1}{2}\right)\nonumber\\
          &- \frac{11}{105} \mathrm{Li}_4\left(\frac{1}{2}\right)
          - \frac{11}{2520} \ln^4(2)
          - \frac{4337}{18900} \ln^5(2)
          - \frac{1411867}{40320} \zeta(5)
          + \frac{4919}{8960} \zeta(4)\nonumber\\
          &+ \frac{280261}{10080} \zeta(4) \ln(2)
          - \frac{1639301}{193536} \zeta(3)
          + \frac{11}{420} \zeta(2) \ln^2(2)
          + \frac{4337}{1890} \zeta(2) \ln^3(2)
          \Big)\nonumber\\
          &+ C_A^2 T_F^2 \Big(
          - \frac{6301303}{65318400} 
          - \frac{8099}{1440} \mathrm{Li}_4\left(\frac{1}{2}\right)
          - \frac{8099}{34560} \ln^4(2)
          + \frac{5}{144} \zeta(5)
          + \frac{30103}{5120} \zeta(4)\nonumber\\
          &- \frac{18564121}{4838400} \zeta(3)
          + \frac{8099}{5760} \zeta(2) \ln^2(2)
          \Big)
          + C_A C_F^2 T_F \Big(
          - \frac{556181}{145152} 
          - \frac{14458}{315} \mathrm{Li}_5\left(\frac{1}{2}\right)\nonumber\\
          &- \frac{39521}{560} \mathrm{Li}_4\left(\frac{1}{2}\right)
          - \frac{39521}{13440} \ln^4(2)
          + \frac{7229}{18900} \ln^5(2)
          + \frac{1214657}{20160} \zeta(5)
          + \frac{3818767}{53760} \zeta(4)\nonumber\\
          &- \frac{13991}{315} \zeta(4) \ln(2)
          - \frac{1990813}{48384} \zeta(3)
          + \frac{39521}{2240} \zeta(2) \ln^2(2)
          - \frac{7229}{1890} \zeta(2) \ln^3(2)
          \Big)\nonumber\nonumber\\
          &+ C_A C_F T_F^2 \Big(
          \frac{12072043}{8164800} 
          + \frac{1457}{90} \mathrm{Li}_4\left(\frac{1}{2}\right)
          + \frac{1457}{2160} \ln^4(2)
          - \frac{5}{24} \zeta(5)
          - \frac{24673}{1440} \zeta(4)\nonumber\\
          &+ \frac{8133593}{806400} \zeta(3)
          - \frac{1457}{360} \zeta(2) \ln^2(2)
          \Big)
          + C_A T_F^3 \Big(
          \frac{6641}{1306368} 
          - \frac{545}{18144} \zeta(3)
          \Big)\nonumber\\
          &+ C_F^3 T_F \Big(
          \frac{37441}{34560} 
          + \frac{256}{15} \mathrm{Li}_5\left(\frac{1}{2}\right)
          + \frac{1919}{45} \mathrm{Li}_4\left(\frac{1}{2}\right)
          + \frac{1919}{1080} \ln^4(2)
          - \frac{32}{225} \ln^5(2)\nonumber\\
          &- \frac{3429}{160} \zeta(5)
          - \frac{58001}{1440} \zeta(4)
          + \frac{212}{15} \zeta(4) \ln(2)
          + \frac{7549}{320} \zeta(3)
          - \frac{1919}{180} \zeta(2) \ln^2(2)\nonumber\\
          &+ \frac{64}{45} \zeta(2) \ln^3(2)
          \Big)
          + C_F^2 T_F^2 \Big(
          \frac{2337647}{1036800} 
          + \frac{874}{45} \mathrm{Li}_4\left(\frac{1}{2}\right)
          + \frac{437}{540} \ln^4(2)
          - \frac{29737}{1440} \zeta(4)\nonumber\\
          &+ \frac{123149}{10800} \zeta(3)
          - \frac{437}{90} \zeta(2) \ln^2(2)
          \Big)
          + C_F T_F^3 \Big(
          - \frac{610843}{3265920} 
          + \frac{661}{3780} \zeta(3)
          \Big)\nonumber\\
          &+ \frac{d_R^{abcd}d_A^{abcd}}{N_A} \Big(
          \frac{6617}{30240}
          + \frac{7496}{105} \mathrm{Li}_5\left(\frac{1}{2}\right)
          + \frac{3988}{105} \mathrm{Li}_4\left(\frac{1}{2}\right)
          + \frac{997}{630} \ln^4(2)
          - \frac{937}{1575} \ln^5(2)\nonumber\\
          &- \frac{274067}{3360} \zeta(5)
          - \frac{194179}{6720} \zeta(4)
          + \frac{49661}{840} \zeta(4) \ln(2)
          + \frac{322631}{20160} \zeta(3)
          - \frac{997}{105} \zeta(2) \ln^2(2)\nonumber\\
          &+ \frac{1874}{315} \zeta(2) \ln^3(2)
          \Big)
          + \frac{d_R^{abcd}d_R^{abcd}}{N_A} \Big(
          - \frac{2411}{5040} 
          + \frac{73}{6} \mathrm{Li}_4\left(\frac{1}{2}\right)
          + \frac{73}{144} \ln^4(2)
          + \frac{5}{12} \zeta(5)\nonumber\\
          &- \frac{2189}{192} \zeta(4)
          + \frac{6779}{1120} \zeta(3)
          - \frac{73}{24} \zeta(2) \ln^2(2)
          \Big)\nonumber\\
          &+ n_l \Bigg[
          C_A^2 T_F^2\Big(
          - \frac{252017}{373248} 
          - \frac{5}{16} \mathrm{Li}_4\left(\frac{1}{2}\right)
          - \frac{5}{384} \ln^4(2)
          + \frac{5}{72} \zeta(5)
          - \frac{59}{512} \zeta(4)\nonumber\\
          &+ \frac{11813}{27648} \zeta(3)
          + \frac{5}{64} \zeta(2) \ln^2(2)
          \Big)
          + C_A C_F T_F^2\Big(
          - \frac{35455}{62208} 
          + \frac{143}{72} \mathrm{Li}_4\left(\frac{1}{2}\right)
          + \frac{143}{1728} \ln^4(2)\nonumber\\
          &- \frac{9359}{2304} \zeta(4)
          + \frac{45287}{13824} \zeta(3)
          - \frac{143}{288} \zeta(2) \ln^2(2)
          \Big)
          + C_A T_F^3\Big(
          \frac{4171}{62208} 
          + \frac{1}{12} \mathrm{Li}_4\left(\frac{1}{2}\right)\nonumber\\
          &+ \frac{1}{288} \ln^4(2)
          - \frac{49}{384} \zeta(4)
          - \frac{59}{3456} \zeta(3)
          - \frac{1}{48} \zeta(2) \ln^2(2)
          \Big)
          + C_F^2 T_F^2\Big(
          - \frac{19}{324} 
          - \frac{49}{18} \mathrm{Li}_4\left(\frac{1}{2}\right)\nonumber\\
          &- \frac{49}{432} \ln^4(2)
          + \frac{1453}{576} \zeta(4)
          - \frac{1955}{1728} \zeta(3)
          + \frac{49}{72} \zeta(2) \ln^2(2)
          \Big)
          + C_F T_F^3\Big(
          - \frac{8663}{93312} \nonumber\\
          &+ \frac{1}{6} \mathrm{Li}_4\left(\frac{1}{2}\right)
          + \frac{1}{144} \ln^4(2)
          - \frac{49}{192} \zeta(4)
          + \frac{77}{432} \zeta(3)
          - \frac{1}{24} \zeta(2) \ln^2(2)
          \Big)\nonumber\\
          &+ \frac{d_R^{abcd}d_R^{abcd}}{N_A}\Big(
          - \frac{103}{216} 
          + \frac{5}{6} \zeta(5)
          + \frac{1}{2}\zeta(4)
          - \frac{131}{72} \zeta(3)
          \Big)
          \Bigg]\nonumber\\
          &+ n_l^2 \Bigg[
          C_A T_F^3\Big(
          - \frac{841}{62208} 
          - \frac{5}{216} \zeta(3)
          \Big)
          + C_F T_F^3\Big(
          - \frac{31147}{93312} 
          + \frac{53}{216} \zeta(3)
          \Big)
          \Bigg]\nonumber\\
          &+ \Bigg[
          C_A^3 T_F \Big(
          - \frac{110041}{124416} 
          + \frac{1577}{9216} \zeta(3)
          \Big)
          + C_A^2 C_F T_F \Big(
          \frac{105763}{20736} 
          - \frac{5105}{1536} \zeta(3)
          \Big)\nonumber\\
          &+ C_A^2 T_F^2 \Big(
          - \frac{2093}{3888}
          + \frac{1}{768} \zeta(3)
          \Big)
          + C_A C_F^2 T_F \Big(
          - \frac{3491}{1152} 
          + \frac{407}{384} \zeta(3)
          \Big)
          + C_A C_F T_F^2 \Big(
          - \frac{8875}{7776}\nonumber\\
          &+ \frac{1963}{1152} \zeta(3)
          \Big)
          + C_A T_F^3 \Big(
          - \frac{437}{7776} 
          + \frac{7}{96} \zeta(3)
          \Big)
          + \frac{157}{128} C_F^3 T_F
          + C_F^2 T_F^2 \Big(
          + \frac{277}{1728} 
          - \frac{67}{144} \zeta(3)
          \Big)\nonumber\\
          &+ C_F T_F^3 \Big(
          - \frac{545}{3888}
          + \frac{7}{48} \zeta(3)
          \Big)
          + \frac{d_R^{abcd}d_A^{abcd}}{N_A} \Big(
          \frac{2}{9} 
          - \frac{13}{6} \zeta(3)
          \Big)\nonumber\\
          &+ \frac{d_R^{abcd}d_R^{abcd}}{N_A} \Big(
          - \frac{11}{36} 
          + \frac{2}{3} \zeta(3)
          \Big)
          \Bigg]\ln\left(\frac{\mu^2}{m^2}\right) + \Bigg[
          - \frac{1993}{6912} C_A^3 T_F
          + \frac{1289}{1728} C_A^2 C_F T_F\nonumber\\
          &+ \frac{1027}{1152} C_A^2 T_F^2
          - \frac{55}{128} C_A C_F^2 T_F
          - \frac{53}{54} C_A C_F T_F^2
          + \frac{49}{864} C_A T_F^3
          + \frac{3}{8} C_F^2 T_F^2\nonumber\\
          &- \frac{17}{144} C_F T_F^3
          \Bigg]\ln^2\left(\frac{\mu^2}{m^2}\right)
          + \Bigg[
          - \frac{77}{1728} C_A^3 T_F
          + \frac{121}{1728} C_A^2 C_F T_F
          + \frac{35}{432} C_A^2 T_F^2\nonumber\\
          &- \frac{55}{432} C_A C_F T_F^2
          - \frac{65}{324} C_A T_F^3
          + \frac{13}{108} C_F T_F^3
          \Bigg]\ln^3\left(\frac{\mu^2}{m^2}\right)
          + \frac{1}{81}T_F^4\ln^4\left(\frac{\mu^2}{m^2}\right)\nonumber\\
          &+ n_l\Bigg[
          C_A^2 T_F^2 \Big(
          \frac{12421}{31104} 
          + \frac{151}{768} \zeta(3)
          \Big)
          + C_A C_F T_F^2 \Big(
          - \frac{9605}{7776} 
          + \frac{1145}{1152} \zeta(3)
          \Big)
          + C_A T_F^3 \Big(
          - \frac{41}{3888} \nonumber\\
          &+ \frac{7}{192} \zeta(3)
          \Big)
          + C_F^2 T_F^2 \Big(
          \frac{73}{864} 
          - \frac{127}{288} \zeta(3)
          \Big)
          + C_F T_F^3 \Big(
          \frac{917}{3888} 
          + \frac{7}{96} \zeta(3)
          \Big)
          + \frac{d_R^{abcd}d_R^{abcd}}{N_A} \Big(
          - \frac{11}{18} \nonumber\\
         &+ \frac{4}{3} \zeta(3)
          \Big)
          \Bigg] \ln\left(\frac{\mu^2}{m^2}\right)
          + n_l^2\Bigg[
          \frac{161}{7776} C_A T_F^3
          + \frac{677}{3888} C_F T_F^3
          \Bigg] \ln\left(\frac{\mu^2}{m^2}\right)
          + n_l\Bigg[
          \frac{55}{1728} C_A^2 T_F^2\nonumber\\
          &- \frac{55}{216} C_A C_F T_F^2
          - \frac{11}{96} C_A T_F^3
          + \frac{11}{48} C_F^2 T_F^2
          - \frac{49}{432} C_F T_F^3
          \Bigg] \ln^2\left(\frac{\mu^2}{m^2}\right)
          + n_l^2\Bigg[
          - \frac{5}{864} C_A T_F^3\nonumber\\
          &- \frac{1}{108} C_F T_F^3
          \Bigg] \ln^2\left(\frac{\mu^2}{m^2}\right)
          + n_l\Bigg[
          \frac{7}{432} C_A^2 T_F^2
          - \frac{11}{144} C_A C_F T_F^2
          + \frac{5}{54} C_F T_F^3
          \Bigg] \ln^3\left(\frac{\mu^2}{m^2}\right)\nonumber\\
          &+ \frac{1}{54} n_l^2 C_F T_F^3\ln^3\left(\frac{\mu^2}{m^2}\right)
~,\\
\zeta_{m}^{(1)} &= 0~,\nonumber\\
\zeta_{m}^{(2)} &= \frac{89}{288} C_F T_F - \frac{5}{24} C_F T_F \ln\left(\frac{\mu^2}{m^2}\right) + \frac{1}{8} C_F T_F \ln^2\left(\frac{\mu^2}{m^2}\right)~,\nonumber\\
\zeta_{m}^{(3)} &= C_A C_F T_F\Big(
          \frac{16627}{15552} 
          - 2 \mathrm{Li}_4\left(\frac{1}{2}\right)
          - \frac{1}{12} \ln^4(2)
          + \frac{31}{16} \zeta(4)
          - \frac{629}{576} \zeta(3)
          + \frac{1}{2} \zeta(2) \ln^2(2)
          \Big)\nonumber\\
          &+ C_F^2 T_F\Big(
          - \frac{683}{576} 
          + 4 \mathrm{Li}_4\left(\frac{1}{2}\right)
          + \frac{1}{6} \ln^4(2)
          - \frac{11}{4} \zeta(4)
          + \frac{57}{32} \zeta(3)
          - \zeta(2) \ln^2(2)
          \Big)\nonumber\\
          &+ C_F T_F^2\Big(
          - \frac{1685}{7776} 
          + \frac{7}{18} \zeta(3)
          \Big)
          + n_l C_F T_F^2\Big(
          \frac{1327}{3888} 
          - \frac{2}{9} \zeta(3)
          \Big)
          + \Bigg[
          C_A C_F T_F\Big(
          \frac{5}{64} 
          - \frac{3}{4} \zeta(3)
          \Big)\nonumber\\
          &+ C_F^2 T_F\Big(
          - \frac{13}{64} 
          + \frac{3}{4} \zeta(3)
          \Big)
          - \frac{31}{108} C_F T_F^2
          \Bigg]\ln\left(\frac{\mu^2}{m^2}\right)
          + \Bigg[
          \frac{29}{96} C_A C_F T_F
          - \frac{1}{4} C_F^2 T_F\nonumber\\
          &+ \frac{5}{72} C_F T_F^2
          \Bigg]\ln^2\left(\frac{\mu^2}{m^2}\right)
          + \Bigg[
          \frac{11}{144} C_A C_F T_F
          - \frac{1}{18} C_F T_F^2
          \Bigg]\ln^3\left(\frac{\mu^2}{m^2}\right)\nonumber\\
          &-\frac{53}{144} n_l C_F T_F^2\ln\left(\frac{\mu^2}{m^2}\right)-\frac{1}{36} n_l C_F T_F^2\ln^3\left(\frac{\mu^2}{m^2}\right)
~,\nonumber\\
\zeta_{m}^{(4)} &= C_A^2 C_F T_F\Big(
          \frac{4524863}{829440} 
          - \frac{173}{15} \mathrm{Li}_5\left(\frac{1}{2}\right)
          - \frac{14539}{640} \mathrm{Li}_4\left(\frac{1}{2}\right)
          - \frac{14539}{15360} \ln^4(2)
          + \frac{173}{1800} \ln^5(2)\nonumber\\
          &- \frac{5}{32} \zeta(6)
          + \frac{7551}{1280} \zeta(5)
          + \frac{759689}{30720} \zeta(4)
          - \frac{1883}{120} \zeta(4) \ln(2)
          - \frac{1640279}{184320} \zeta(3)
          - \frac{21}{128} \zeta(3)^2\nonumber\\
          &+ \frac{14539}{2560} \zeta(2) \ln^2(2)
          - \frac{173}{180} \zeta(2) \ln^3(2)
          \Big)
          + C_A C_F^2 T_F\Big(
          \frac{1068103}{414720} 
          + \frac{514}{15} \mathrm{Li}_5\left(\frac{1}{2}\right)\nonumber\\
          &+ \frac{11321}{320} \mathrm{Li}_4\left(\frac{1}{2}\right)
          + \frac{11321}{7680} \ln^4(2)
          - \frac{257}{900} \ln^5(2)
          - \frac{425}{128} \zeta(6)
          - \frac{77977}{1920} \zeta(5)
          - \frac{181317}{5120} \zeta(4)\nonumber\\
          &+ \frac{1321}{30} \zeta(4) \ln(2)
          + \frac{398489}{30720} \zeta(3)
          - \frac{11}{32} \zeta(3)^2
          - \frac{11321}{1280} \zeta(2) \ln^2(2)
          + \frac{257}{90} \zeta(2) \ln^3(2)
          \Big)\nonumber\\
          &+ C_A C_F T_F^2\Big(
          - \frac{214882117}{203212800} 
          + \frac{28657}{3360} \mathrm{Li}_4\left(\frac{1}{2}\right)
          + \frac{28657}{80640} \ln^4(2)
          + \frac{97}{24} \zeta(5)
          - \frac{152979}{17920} \zeta(4)\nonumber\\
          &+ \frac{29927237}{11289600} \zeta(3)
          - \frac{28657}{13440} \zeta(2) \ln^2(2)
          \Big)
          + C_F^3 T_F\Big(
          \frac{10301}{10240} 
          - \frac{112}{5} \mathrm{Li}_5\left(\frac{1}{2}\right)
          + \frac{3227}{240} \mathrm{Li}_4\left(\frac{1}{2}\right)\nonumber\\
          &+ \frac{3227}{5760} \ln^4(2)
          + \frac{14}{75} \ln^5(2)
          + \frac{65}{64} \zeta(6)
          + \frac{10003}{320} \zeta(5)
          - \frac{20897}{1920} \zeta(4)
          - \frac{253}{10} \zeta(4) \ln(2)\nonumber\\
          &+ \frac{1427}{480} \zeta(3)
          + \frac{87}{32} \zeta(3)^2
          - \frac{3227}{960} \zeta(2) \ln^2(2)
          - \frac{28}{15} \zeta(2) \ln^3(2)
          \Big)
          + C_F^2 T_F^2\Big(
          \frac{257128337}{203212800} \nonumber\\
          &+ \frac{5041}{1680} \mathrm{Li}_4\left(\frac{1}{2}\right)
          + \frac{5041}{40320} \ln^4(2)
          - \frac{63}{16} \zeta(5)
          - \frac{90269}{26880} \zeta(4)
          + \frac{7671973}{1881600} \zeta(3)\nonumber\\
          &- \frac{5041}{6720} \zeta(2) \ln^2(2)
          \Big)
          + C_F T_F^3\Big(
          \frac{1281821}{19595520} 
          + \frac{1}{48} \zeta(4)
          + \frac{51}{560} \zeta(3)
          \Big)
          + \frac{d_R^{abcd}d_R^{abcd}}{N_F}\Big(
          - \frac{611}{384} \nonumber\\
          &+ 40 \mathrm{Li}_4\left(\frac{1}{2}\right)
          + \frac{5}{3} \ln^4(2)
          - \frac{15}{4} \zeta(6)
          - \frac{135}{32} \zeta(5)
          - \frac{1445}{64} \zeta(4)
          + \frac{973}{64} \zeta(3)
          + \frac{15}{4} \zeta(3)^2\nonumber\\
          &- 10 \zeta(2) \ln^2(2)
          \Big)
          + n_l \Bigg[
          C_A C_F T_F^2 \Big(
          - \frac{5095}{3072} 
          + 4 \mathrm{Li}_5\left(\frac{1}{2}\right)
          + \frac{49}{12} \mathrm{Li}_4\left(\frac{1}{2}\right)
          + \frac{49}{288} \ln^4(2)\nonumber\\
          &- \frac{1}{30} \ln^5(2)
          - \frac{253}{96} \zeta(5)
          - \frac{543}{128} \zeta(4)
          + \frac{49}{8} \zeta(4) \ln(2)
          - \frac{65}{192} \zeta(3)
          - \frac{49}{48} \zeta(2) \ln^2(2)\nonumber\\
          &+ \frac{1}{3} \zeta(2) \ln^3(2)
          \Big)
          + C_F^2 T_F^2 \Big(
          - \frac{15557}{5184} 
          - 8 \mathrm{Li}_5\left(\frac{1}{2}\right)
          - \frac{49}{6} \mathrm{Li}_4\left(\frac{1}{2}\right)
          - \frac{49}{144} \ln^4(2)\nonumber\\
          &+ \frac{1}{15} \ln^5(2)
          + \frac{157}{16} \zeta(5)
          + \frac{1639}{192} \zeta(4)
          - \frac{49}{4} \zeta(4) \ln(2)
          - \frac{5}{8} \zeta(3)
          + \frac{49}{24} \zeta(2) \ln^2(2)\nonumber\\
          &- \frac{2}{3} \zeta(2) \ln^3(2)
          \Big)
          + C_F T_F^3 \Big(
          - \frac{57}{256} 
          + \frac{9}{16} \zeta(4)
          - \frac{5}{144} \zeta(3)
          \Big)
          \Bigg]
          + n_l^2 C_F T_F^3 \Big(
          \frac{17671}{20736} 
          - \frac{7}{16} \zeta(4)\nonumber\\
          &- \frac{5}{144} \zeta(3)
          \Big)
          + \Bigg[
          C_A^2 C_F T_F \Big(
          \frac{233903}{248832} 
          - \frac{11}{2} \mathrm{Li}_4\left(\frac{1}{2}\right)
          - \frac{11}{48} \ln^4(2)
          + \frac{25}{16} \zeta(5)
          + \frac{407}{64} \zeta(4)\nonumber\\
          &- \frac{119723}{18432} \zeta(3)
          + \frac{11}{8} \zeta(2) \ln^2(2)
          \Big)
          + C_A C_F^2 T_F \Big(
          - \frac{3529}{768} 
          + 11 \mathrm{Li}_4\left(\frac{1}{2}\right)
          + \frac{11}{24} \ln^4(2)\nonumber\\
          &+ \frac{5}{16} \zeta(5)
          - \frac{275}{32} \zeta(4)
          + \frac{8913}{1024} \zeta(3)
          - \frac{11}{4} \zeta(2) \ln^2(2)
          \Big)
          + C_A C_F T_F^2 \Big(
          - \frac{39259}{20736} \nonumber\\
          &+ 2 \mathrm{Li}_4\left(\frac{1}{2}\right)
          + \frac{1}{12} \ln^4(2)
          - \frac{37}{16} \zeta(4)
          + \frac{4343}{1536} \zeta(3)
          - \frac{1}{2} \zeta(2) \ln^2(2)
          \Big)
          + C_F^3 T_F \Big(
          \frac{217}{768} \nonumber\\
          &- \frac{15}{8} \zeta(5)
          + \frac{169}{256} \zeta(3)
          \Big)
          + C_F^2 T_F^2 \Big(
          \frac{8951}{6912} 
          - 4 \mathrm{Li}_4\left(\frac{1}{2}\right)
          - \frac{1}{6} \ln^4(2)
          + \frac{25}{8} \zeta(4)
          - \frac{595}{256} \zeta(3)\nonumber\\
          &+ \zeta(2) \ln^2(2)
          \Big)
          + C_F T_F^3 \Big(
          \frac{359}{1944} 
          - \frac{1}{3} \zeta(3)
          \Big)
          + \frac{d_R^{abcd}d_R^{abcd}}{N_F} \Big(
          \frac{1}{4} 
          - \frac{15}{8} \zeta(3)
          \Big)
          \Bigg]\ln\left(\frac{\mu^2}{m^2}\right)\nonumber\\
          &+ \Bigg[
          C_A^2 C_F T_F \Big(
          \frac{19867}{13824} 
          - \frac{33}{32} \zeta(3)
          \Big)
          + C_A C_F^2 T_F \Big(
          - \frac{219}{128} 
          + \frac{33}{32} \zeta(3)
          \Big)
          + C_A C_F T_F^2 \Big(
          - \frac{2059}{6912} \nonumber\\
          &+ \frac{3}{8} \zeta(3)
          \Big)
          + \frac{15}{16} C_F^3 T_F
          + C_F^2 T_F^2 \Big(
          \frac{193}{2304} 
          - \frac{3}{8} \zeta(3)
          \Big)
          + \frac{31}{216} C_F T_F^3
          \Bigg]\ln^2\left(\frac{\mu^2}{m^2}\right)\nonumber\\
          &+ \Bigg[
          \frac{17}{48} C_A^2 C_F T_F
          - \frac{143}{384} C_A C_F^2 T_F
          - \frac{13}{48} C_A C_F T_F^2
          + \frac{31}{192} C_F^2 T_F^2
          - \frac{5}{216} C_F T_F^3
          \Bigg]\ln^3\left(\frac{\mu^2}{m^2}\right)\nonumber\\
          &+ \Bigg[
          \frac{121}{2304} C_A^2 C_F T_F
          - \frac{11}{192} C_A C_F T_F^2
          + \frac{1}{128} C_F^2 T_F^2
          + \frac{1}{48} C_F T_F^3
          \Bigg]\ln^4\left(\frac{\mu^2}{m^2}\right)\nonumber\\
          &+ n_l\Bigg[
          C_A C_F T_F^2 \Big(
          - \frac{5155}{31104} 
          + 2 \mathrm{Li}_4\left(\frac{1}{2}\right)
          + \frac{1}{12} \ln^4(2)
          - \frac{43}{16} \zeta(4)
          + \frac{997}{576} \zeta(3)\nonumber\\
          &- \frac{1}{2} \zeta(2) \ln^2(2)
          \Big)\nonumber
          + C_F^2 T_F^2 \Big(
          \frac{319}{192} 
          - 4 \mathrm{Li}_4\left(\frac{1}{2}\right)
          - \frac{1}{6} \ln^4(2)
          + \frac{7}{2} \zeta(4)
          - \frac{97}{32} \zeta(3)\\
          &+ \zeta(2) \ln^2(2)
          \Big)
          - \frac{143}{648} C_F T_F^3
          \Bigg]\ln\left(\frac{\mu^2}{m^2}\right)
          + n_l^2 C_F T_F^3\Big(
          - \frac{3401}{7776} 
          + \frac{7}{18} \zeta(3)
          \Big)\ln\left(\frac{\mu^2}{m^2}\right)\nonumber\\
          &+ n_l\Bigg[
          - \frac{2581}{3456} C_A C_F T_F^2
          - \frac{9}{64} C_F^2 T_F^2
          + \frac{283}{864} C_F T_F^3
          \Bigg]\ln^2\left(\frac{\mu^2}{m^2}\right)
          + \frac{31}{216} n_l^2 C_F T_F^3\ln^2\left(\frac{\mu^2}{m^2}\right)\nonumber\\
          &+ n_l\Bigg[
          - \frac{13}{96} C_A C_F T_F^2
          + \frac{1}{8} C_F^2 T_F^2
          \Bigg]\ln^3\left(\frac{\mu^2}{m^2}\right)
          + n_l\Bigg[
          - \frac{11}{288} C_A C_F T_F^2\nonumber\\
          &+ \frac{1}{48}  C_F T_F^3
          \Bigg]\ln^4\left(\frac{\mu^2}{m^2}\right)
          + \frac{1}{144} n_l^2 C_F T_F^3\ln^4\left(\frac{\mu^2}{m^2}\right)
~.
\end{align}
and
\begin{align}
\zeta_{1}^{(2)} &= \frac{5}{96} C_F T_F + \frac{89}{1152} C_A T_F - \left(\frac{1}{8} C_F T_F + \frac{5}{96}C_A T_F\right)\ln\left(\frac{\mu^2}{m^2}\right) + \frac{1}{32} C_A T_F\ln^2\left(\frac{\mu^2}{m^2}\right)~,\\
\zeta_{2}^{(2)} &= \frac{5}{96} C_F T_F - \frac{1}{8} C_F T_F \ln\left(\frac{\mu^2}{m^2}\right)~,\\
\zeta_{3}^{(1)} &= - \frac{1}{3} T_F \ln\left(\frac{\mu^2}{m^2}\right)~,\\
\tilde{\zeta}_{1}^{(3)} &= \left(1-\xi^{(n_f)}\right)\Big(C_A^2 T_F\left(\frac{2039}{27648} - \frac{1}{48}\zeta(3)\right) - \frac{7}{144} C_A^2 T_F \ln\left(\frac{\mu^2}{m^2}\right)\nonumber\\
          & + \frac{5}{384} C_A^2 T_F\ln^2\left(\frac{\mu^2}{m^2}\right) - \frac{1}{384} C_A^2 T_F\ln^3\left(\frac{\mu^2}{m^2}\right)\Big)~,\\
\tilde{\zeta}_{3}^{(1)} &= - \frac{89}{1152} C_A T_F + \frac{5}{96} C_A T_F \ln\left(\frac{\mu^2}{m^2}\right) - \frac{1}{32} C_A T_F\ln^2\left(\frac{\mu^2}{m^2}\right)~,\\
\zeta_{3g}^{(1)} &= \frac{1}{3} T_F \ln\left(\frac{\mu^2}{m^2}\right)~,\\
\zeta_{4g}^{(1)} &= \frac{1}{3} T_F \ln\left(\frac{\mu^2}{m^2}\right)~.
\end{align}
In these expression $m\equiv m(\mu)$ is the
$\overline{\rm MS}$ quark mass and $N_F = N_c$. Other variants with $\alpha_s^{(n_l)}$ and the
on-shell heavy quark mass can be found in~\cite{progdata}.
The colour factors are defined in Eq.~(\ref{eq::cf}).

\end{appendix}

%- }}}
%- {{{ bibliography:

%- }}}

\end{document}